\documentclass[nopreprintline,12pt]{elsarticle}

\usepackage[a4paper, portrait, margin=1.1811in]{geometry}
\usepackage[english]{babel}
\usepackage[utf8]{inputenc}
\usepackage[T1]{fontenc}
\usepackage{CJK} 
\usepackage{helvet}
\usepackage{etoolbox}
\usepackage{graphicx}
\usepackage{titlesec}
\usepackage{multirow}
\usepackage{ulem} 
\usepackage{makecell}
\usepackage{lineno}
\titleformat{\paragraph}[runin]{\normalfont\bfseries}{\theparagraph}{1em}{}

\usepackage{adjustbox}
\usepackage{xcolor}
\usepackage{array}
\usepackage{graphicx}

\usepackage{algorithm}
\usepackage{algpseudocode}
\algrenewcommand\algorithmicrequire{\textbf{Require:}}
\algrenewcommand\algorithmicensure{\textbf{Ensure:}}
\algrenewcommand\algorithmiccomment[1]{\hfill$\triangleright$ #1}
\usepackage{amsmath, amssymb}

\usepackage{url}
\usepackage{booktabs} 
\usepackage{xcolor} 
\usepackage[colorlinks, citecolor=cyan]{hyperref}
\usepackage{float}
\usepackage{caption}
\usepackage{xurl}
\graphicspath{ {./images/} }
\usepackage{scrextend}
\usepackage{threeparttable}
\usepackage{ragged2e}
\usepackage{fancyhdr}
\usepackage{lipsum,float}
\newcounter{lemma}

\newcounter{theorem}

\usepackage{tabularx}
\newcolumntype{C}[1]{>{\centering\arraybackslash}p{#1}}
\usepackage{amsmath}
\usepackage{subcaption}
\usepackage{hyperref}
\usepackage{longtable}


\journal{Applied Soft Computing}

\begin{document}
\begin{sloppypar}
\begin{frontmatter}



\title{MMformer with Adaptive Attention: Advancing Multivariate Time Series Forecasting for Environmental Applications}


\author[inst1]{Ning Xin}
\ead{Ning.Xin21@student.xjtlu.edu.cn}
\affiliation[inst1]{organization={School of AI and Advanced Computing},
            addressline={XJTLU Entrepreneur College (Taicang)},
            addressline={Xi'an Jiaotong-Liverpool University}, 
            city={Suzhou},
            postcode={215123},
            state={Jiangsu},
            country={China}}      
\author[inst1]{Jionglong Su}
\ead{Jionglong.Su@xjtlu.edu.cn}
\author[inst1]{Md Maruf Hasan \corref{cor1}}
\ead{MdMaruf.Hasan@xjtlu.edu.cn}
            
\cortext[cor1]{Corresponding author}

\begin{abstract}
Multivariate time series forecasting is critical in numerous scientific and engineering fields. Current deep learning models, however, face fundamental challenges in this domain. They struggle to handle dynamically changing data distributions, achieve model self-adaptation, and quantify prediction uncertainty. These limitations are particularly severe in fields like environmental science, where high-reliability decisions are essential. To address these limitations, we introduce MMformer, an innovative common adaptive multivariate time series forecasting model. MMformer's core lies in the synergistic integration of three key components: an encoder-only optimized architecture for capturing temporal dependencies; the Adaptive Transferable Multi-Head Attention mechanism, using meta-learning to enhance model adaptability and generalization; and Monte Carlo Dropout for improved robustness and crucial uncertainty quantification.

We rigorously evaluate MMformer on real-world environmental datasets (e.g., 1277 days of air quality data across 331 Chinese cities, 1826 days of temperature and rainfall data from 909 stations) and general MTS benchmarks (PEMS). Results demonstrate that MMformer achieves superior performance in accuracy, adaptability, and uncertainty quantification than iTransformer, PatchTST, TimesNet, and Transformer. Specifically, on air quality data, MMformer reduced MSE, MAE, and MAPE by up to 70.195\%, 37.240\%, and 36.542\%, respectively, compared to iTransformer; up to 69.641\%, 37.121\%, and 35.632\% compared to TimesNet; up to 71.578\%, 40.288\%, and 39.383\% compared to PatchTST; and up to 68.182\%, 36.641\%, and 35.343\% compared to Transformer. On climate data, the performance improvements mirror those on the air quality dataset. MMformer also achieved optimal results on PEMS03, PEMS04, and PEMS08, demonstrating strong generalization. Our approach offers key theoretical insights for researchers and significant practical value for policymakers. It marks a significant advance in forecasting and managing dynamic environmental challenges. This capability enables precise interventions to protect public health and promote environmental sustainability.
\end{abstract}

\begin{graphicalabstract}
\includegraphics[width=1\textwidth]{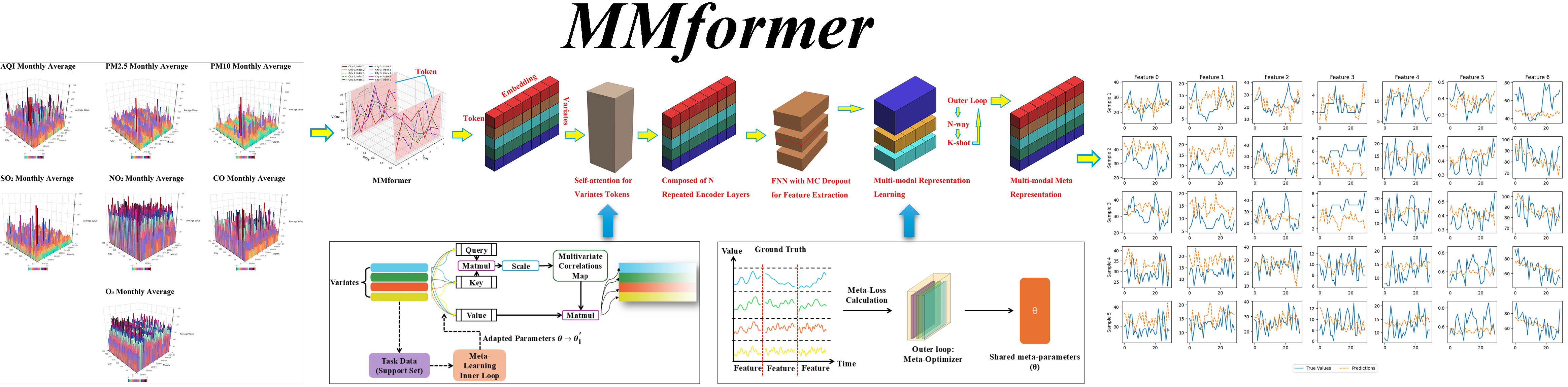}
\centering
\label{graphic abstract}
\end{graphicalabstract}

\begin{highlights}
\item Research Highlight 1

A novel MMformer model with a novel adaptive forecasting framework for multivariate time series prediction.
\item Research Highlight 2

Combining self-attention with meta-learning improves the model's generalization ability and adaptability to different data distributions.
\item Research Highlight 3

MMformer achieves state-of-the-art performance on widely-used MTS benchmarks and challenging real-world environmental prediction tasks, outperforming existing models such as iTransformer, PatchTST, TimesNet, and Transformer.
\end{highlights}

\end{frontmatter}
\section{Introduction}
Deep learning (DL) leads multivariate time series (MTS) forecasting, propelled by machine learning advancements and broad interdisciplinary adoption. Current advanced models, however, face significant limitations. They struggle with dynamic data distributions, model self-adaptation, and quantifying prediction uncertainty \citep{zou2019complex, hewamalage2021recurrent, ismail2020benchmarking, jiang2021applications}. These challenges are particularly critical in high-risk decision-making domains like environmental science \citep{xu2019hybrid,barandas2020tsfel}. Forecast reliability directly impacts public health and sustainable development strategies \citep{sokhi2021advances}. Reliable uncertainty estimation is crucial for effective environmental decision-making  \citep{bressane2024fuzzy}. Standard deep learning architectures frequently lack this vital capability. This study employs advanced soft computing techniques to address these pervasive MTS prediction challenges. We validate their effectiveness in demanding environmental applications.

MTS analysis has evolved from traditional statistical models to DL \citep{moskovitch2022multivariate}. Traditional models, like Generalized Linear Models (GLMs) \citep{8808897}, Seasonal Autoregressive Integrated Moving Average (SARIMA) \citep{en14196021}, and Autoregressive Integrated Moving Average (ARIMA) \citep{8170657}, offer interpretability but struggle with the inherent complexity, nonlinearity, and high dimensionality of modern MTS data. Conversely, DL methods excel at automatic feature extraction and modeling complex data \citep{10.1371/journal.pone.0254841}. The above motivates the development of our innovative DL model with three motivations: (1) enhancing the utility of the predictions, particularly in environmental, through developing robust models and uncertainty analysis; (2) improving prediction reliability and informing environmental decision-making through uncertainty quantification; and (3) improving predictive accuracy, optimizing resource management, and facilitating the understanding of environmental changes by effectively handling spatiotemporal data. However, three key challenges persist: (1) standard DL models can be computationally intensive, less interpretable, and lack uncertainty analysis \citep{9369420}; (2) achieving the adaptability and robustness for reliable uncertainty quantification across diverse real-world conditions remains challenging \citep{he2023survey}; and (3) the predictive performance of existing DL models on spatiotemporal time series is often suboptimal \citep{su2010online}. A performance bottleneck persists for common DL models, which have limited accuracy with complex spatiotemporal dependencies and inter-variable interactions in real-world data \citep{su2010online1}. These persistent challenges necessitate an innovative DL model to enhance MTS prediction adaptability, robustness, and reliability systematically.

To overcome these challenges, this study introduces MMformer, a novel adaptive integrated forecasting model. The core innovation of MMformer lies in the synergistic integration of three key components within a neural network architecture, specifically designed to enhance adaptability and provide uncertainty quantification. These components are: (1) an Encoder-only inverse embedding neural network for capturing temporal dependencies \citep{liu2023itransformer}; (2) an Adaptive Transferable Multi-head Attention (ATMA) mechanism that incorporates meta-learning strategies \citep{electronics12040837}, specifically chosen to significantly enhance the model's adaptability and generalization across multiple regions and data distributions, thereby addressing the need for domain-specific solutions; and (3) Monte Carlo Dropout (MC Dropout) \citep{laves2019wellcalibrated}, strategically applied to improve regularization, model robustness against noise, and crucially, to provide vital uncertainty quantification for better risk assessment. The integration of meta-learning addresses the need for effective adaptation, and MC Dropout enhances interpretability through uncertainty modeling, directly responding to identified motivations for model design.

We evaluate MMformer's performance against state-of-the-art (SOTA) and baseline models, including iTransformer \citep{liu2023itransformer}, PatchTST \citep{zhang2024patchtcn}, TimesNet \citep{wu2023timesnettemporal2dvariationmodeling}, and Transformer \citep{ijcai2023p759}, across diverse datasets. The evaluation encompasses the PEMS datasets \citep{frbe-bw46-20}, and two challenged real-world environmental datasets: (1) seven major pollutants across 331 Chinese cities (January 2018 - June 2021, 1277 days) \citep{horn_air_2024}, and (2) precipitation and temperature data from 909 sites (2018-2022, 1826 days) \citep{cmdc}. Validation on these real-world datasets, alongside the widely used MTS dataset, addresses concerns about domain specificity and dataset validity. The comprehensive comparisons demonstrate significant performance gains, proving the model's effectiveness.

The contributions of this work are as follows:
\begin{itemize}
    \item[1.] We introduce MMformer, an innovative model designed for the specific challenges of MTS forecasting across multiple regions. By integrating an encoder-only network, the ATMA module, and MC Dropout, MMformer effectively captures complex time features, generalizes across diverse conditions, and provides robust uncertainty quantification. The design directly addresses the limitations of DL models in risk assessment and decision-making.

    \item[2.] We developed the ATMA module, which optimizes the attention mechanism and improves overall model performance. ATMA enables rapid adaptation to different regions' prediction tasks, enhances data heterogeneity and dynamics robustness, and reduces reliance on domain-specific knowledge. By performing gradient updates on each sub-task, ATMA facilitates fast adaptation and improves generalization across tasks via meta-gradients, thereby contributing to overall model performance and generalizing across different data distributions and varying regions.

    \item[3.] We incorporate MC Dropout to enhance model robustness in noisy environments and provide prediction uncertainty quantification, enhancing result interpretability and supporting risk assessment and decision-making.

    \item[4.] Empirical results demonstrate MMformer's superior performance in MTS prediction, particularly in forecasting for multiple regions. On air quality datasets, MMformer achieves significant improvements, reducing MSE by 68.182\%–71.578\%, MAE by 36.641\%–40.288\%, and MAPE by 35.343\%–39.383\% compared to iTransformer, Transformer, TimesNet, and PatchTST. MMformer also exhibited excellent performance on climate datasets, with all evaluation metrics decreasing by 27.837\%–55.363\% relative to baselines. Experiments on the PEMS datasets further validate MMformer's generalization capability. On PEMS03, MMformer achieves the best performance, with all evaluation metrics decreasing by 3.352\%–43.103\% compared to baselines. On PEMS04, MMformer also demonstrated the best performance, reducing all evaluation metrics by 10.714\%-39.458\%. On PEMS07, MMformer's MSE is comparable to TimesNet but underperformed iTransformer; however, it still improves all metrics compared to PatchTST and Transformer, reducing by 7.143\%-54.819\%. On PEMS08, MMformer shows the best performance, reducing all evaluation metrics by 10.672\% - 69.470\%. The above experimental outcomes underscore MMformer's strong generalization ability and robustness in complex multiple regions MTS forecasting.
\end{itemize}

The structure of this paper is organized as follows: Section 2 reviews MTS forecasting methods. Section 3 details the proposed approach. Section 4 presents the experimental design and results. Section 5 discusses the research objectives and the research results in relation to previous work. Finally, Section 6 concludes and discusses future work.

\section{Related Work}
MTS forecasting is critical in fields such as environmental science. Methodologies have advanced significantly with the rapid development of neural network architectures, notably Transformers and their variants \citep{10.1145/3447548.3467401, HU2022109092, 9964035, kim_feat_2023, foumani_improving_2023}. While traditional statistical methods (e.g., ARIMA, SARIMA) and early machine learning techniques (e.g., VAR \citep{WOS:000570153300001}, SVMs \citep{bansal2022comparative}, Multi-Layer Perceptrons (MLPs) \citep{WOS:000658511600006}) once dominated time series analysis, they exhibit limitations when addressing the inherent complexities of real-world MTS data. These limitations include handling high dimensionality, intricate nonlinear relationships, spatio-temporal dynamics, cross-domain adaptability, and capturing subtle inter-series dependencies and long-range dependencies. Consequently, research focus has shifted towards DL models to more effectively tackle these key MTS forecasting challenges. This section will detail the evolution of DL in MTS analysis, emphasizing how these models overcome challenges and discussing current progress in model robustness and generalization, thereby laying the theoretical groundwork for the proposed MMformer.

\subsection{Early Deep Learning Models for MTS Forecasting: RNNs and LSTMs}
Deep learning has driven the evolution of MTS forecasting, transitioning from traditional statistical and early machine learning methods to neural network-based approaches. The shift is underpinned by deep learning's superior capacity to capture intrinsic temporal dependencies \citep{WOS:000829197200001, bi2023hierarchical} and its reduction in the need for extensive manual feature engineering \citep{9461796, WEN2021167, 9075431}. Recurrent Neural Networks (RNNs) \citep{KAMILARIS201870}, including variants like Long Short-Term Memory (LSTM) \citep{zhao_remaining_2017} and Gated Recurrent Unit (GRU) \citep{shu2021short}, significantly advanced MTS prediction by modeling temporal dependencies. RNNs introduced the capability for information retention across time steps \citep{zargar2021introduction}, while LSTMs and GRUs mitigated the vanishing gradient problem, enabling the capture of long-range dependencies \citep{al2023lstm, zargar2021introduction}. However, RNN-based models exhibit inherent limitations. Their sequential processing architecture hinders parallel computation efficiency for long sequences, leading to increased training times and resource consumption \citep{hewamalage2021recurrent}. They also face challenges in capturing complex inter-variable interactions \citep{hewamalage2021recurrent} and, due to their serial nature, are difficult to parallelize, thereby limiting scalability for large datasets and real-time applications \citep{gonzalez2024towards, zhang2020achieving}.

\subsection{Beyond RNNs: Attention Mechanisms and Transformer Models in MTS Prediction}
Attention mechanisms represent a significant advancement in MTS forecasting, overcoming RNN limitations by enabling models to selectively weight input sequence elements and thus better capture long-range dependencies \citep{soydaner2022attention}. The Transformer architecture, built upon self-attention, directly computes dependencies between any two sequence positions. Attention eliminates reliance on sequential order, greatly enhancing the ability to capture long-range temporal dependencies \citep{ijcai2023p759}. Furthermore, self-attention's inherent parallelizability is crucial for processing high-dimensional, long-span MTS data.

Transformer architectures and their variants have been extensively developed for MTS forecasting, enhancing their capability to model complex dynamics \citep{10.1145/3447548.3467401}. Informer reduced computational complexity via Probsparse self-attention and Auto-correlation mechanisms, facilitating efficient processing of ultra-long sequences \citep{zhou2021informer}. PatchTST segments time series into patches, applying Transformers to capture both local intra-patch and global inter-patch dependencies, achieving strong benchmark performance \citep{zhang2024patchtcn}. TimesNet focuses on time series decomposition, separating trend and seasonality to better model temporal patterns across various scales, resulting in SOTA performance \citep{wu2023timesnettemporal2dvariationmodeling}. Notably, iTransformer reorients attention to the variable dimension, directly addressing the critical task of capturing inter-variable correlations in MTS \citep{liu2023itransformer}.

These attention-based and Transformer models excel at capturing complex nonlinearities, spatio-temporal dynamics, and subtle inter-variable dependencies in MTS data through their robust feature extraction and deep sequence modeling \citep{ijcai2023p759}. iTransformer's specific focus on inter-variable correlations directly addresses a core MTS challenge. Nevertheless, standard Transformers retain a quadratic computational complexity issue when processing extremely long sequences. Ongoing research thus focuses on novel optimization strategies to improve precision, robustness, and adaptability to diverse data scales and heterogeneities \citep{li_shapenet_2021, rs12101688}.

\subsection{Uncertainty Quantification, Adaptability, and Meta-Learning in Deep Learning for MTS Forecasting}
While Transformers have significantly advanced MTS forecasting, particularly in handling long-range dependencies and enabling parallel processing, the complexity of real-world MTS data, especially in applications across multiple regions, necessitates further exploration \citep{shih_temporal_2019, lubba_catch22_2019}. It includes capturing finer aspects of data dynamics, such as inherent predictive uncertainty and model adaptability, crucial for reliable forecasts and informed decision-making \citep{wu2021ensemble}.

Probabilistic forecasting and uncertainty quantification are vital for high-stakes applications like environmental monitoring, where understanding forecast uncertainty is as important as the prediction itself \citep{wei2021uncertainty}. Advanced deep learning methods provide mechanisms to quantify uncertainty beyond point predictions. Techniques like MC Dropout and Bayesian neural networks enable models to express confidence levels in their forecasts \citep{laves2019wellcalibrated, nguyen2023long}. MC Dropout, in particular, offers a practical approach to estimate prediction variance, a capability that deterministic models typically lack. Uncertainty quantification is essential for risk assessment and decision-making in dynamic environments.

Real-world MTS datasets often exhibit significant statistical property variations across different tasks or time periods \citep{shao2024exploring}. Meta-learning and transfer learning have emerged as key solutions to enhance model adaptability and generalization \citep{khoee2024domain}. These methods enable models to "learn to learn" by acquiring knowledge from task distributions. It facilitates rapid adaptation to novel or unseen MTS forecasting challenges, even with limited target task data \citep{vettoruzzo2024advances}. The focus on rapid adaptation directly addresses the heterogeneity prevalent in environmental data.

Through these advanced deep learning techniques, researchers are developing more robust, adaptable, and interpretable MTS forecasting models to address the complex challenges posed by real-world spatiotemporal time series data.

\subsection{MMformer: A Meta-Learning Framework for Adaptive MTS Forecasting}
MTS prediction, particularly in environmental science, faces critical challenges in model adaptability, cross-task generalization, and quantifying prediction uncertainty. While Transformer architectures advance long-range dependency capture and parallel computation, standard models are limited by high-dimensional MTS data, complex inter-variable relationships, and data heterogeneity. To address these, MMformer is proposed, building on prior work, notably iTransformer's dimension inversion approach, to resolve key MTS prediction challenges.

MMformer's core innovation is its ATMA mechanism, integrating Meta-Learning for rapid parameter adaptation and optimization when facing data distribution shifts or new tasks. The capability addresses the critical need for domain adaptability across multiple regions of MTS data with varying conditions. Furthermore, MMformer's embedding strategy, which has timestamps and transposes dimensions, enhances its capture of intrinsic spatio-temporal dynamics.

To bolster prediction reliability, MMformer strategically incorporates MC Dropout. MC Dropout enhances robustness against noise and quantifies uncertainty, vital for risk assessment and decision-making where deterministic models fall short. By effectively capturing diverse data patterns and leveraging its structural diversity, MMformer improves prediction accuracy, scalability, and robustness for real-time applications. These integrated innovations offer a more robust and intelligent solution for complex multi-regional MTS forecasting, overcoming previous limitations.

\section{Methodology}
In this section, we shall first present the data preprocessing methods and then comprehensively describe the methodology, including detailed procedures and clear flowcharts. It is followed by the model's components, detailing the innovative components. These advancements lead to substantial improvements in neural network optimization and increase the accuracy of time-series forecasting.

\subsection{Data Preprocessing}
This study used a comprehensive data preprocessing step to ensure the quality and consistency of the MTS data input to the model. Fig.\ref{fig1} is the flowchart of data preprocessing, and the preprocessing process includes the following key steps:
\begin{figure}[H]
    \centering
    \includegraphics[width=0.9\textwidth]{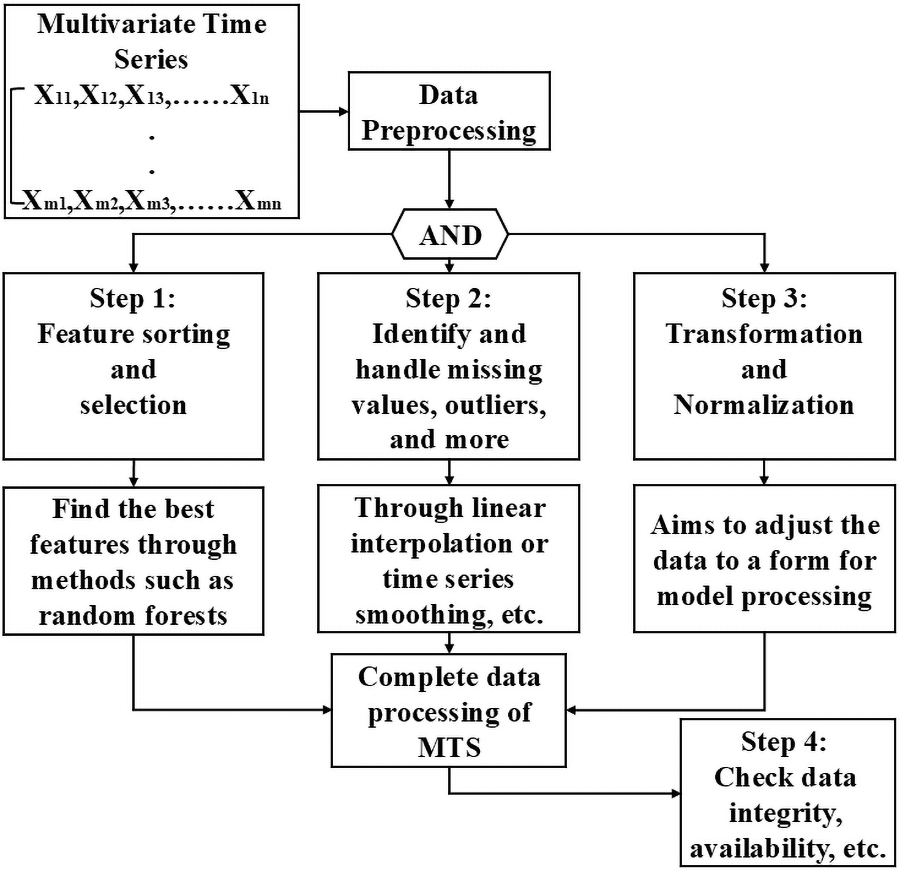}
    \caption{Flowchart of Data Preprocessing}
    \label{fig1}
\end{figure}

\begin{itemize}
\item
\textbf{Step 1. Feature sorting and selection:}
Features are sorted and selected, aiming to identify the prediction tasks with the most relevant and influential features. This step helps to reduce noise and improve the robustness and accuracy of the model.

\item
\textbf{Step 2. Missing value and outlier processing:}
Missing values and outliers in the dataset are treated using techniques such as linear interpolation or MTS smoothing to ensure the continuity and reliability of the data.

\item
\textbf{Step 3. Data transformation and normalization:}
The selected features are subjected to necessary data transformation and normalization. This step aims to adjust the data to a form suitable for model processing.

\item
\textbf{Step 4. Integrity check:}
In the final stage of preprocessing, the processed data is checked for integrity, including verification of data integrity, availability, etc., to ensure that the data quality meets the requirements of model training.
\end{itemize}

\subsection{Workflow Overview}
\begin{figure}[ht]
    \centering
    \includegraphics[width=1\textwidth]{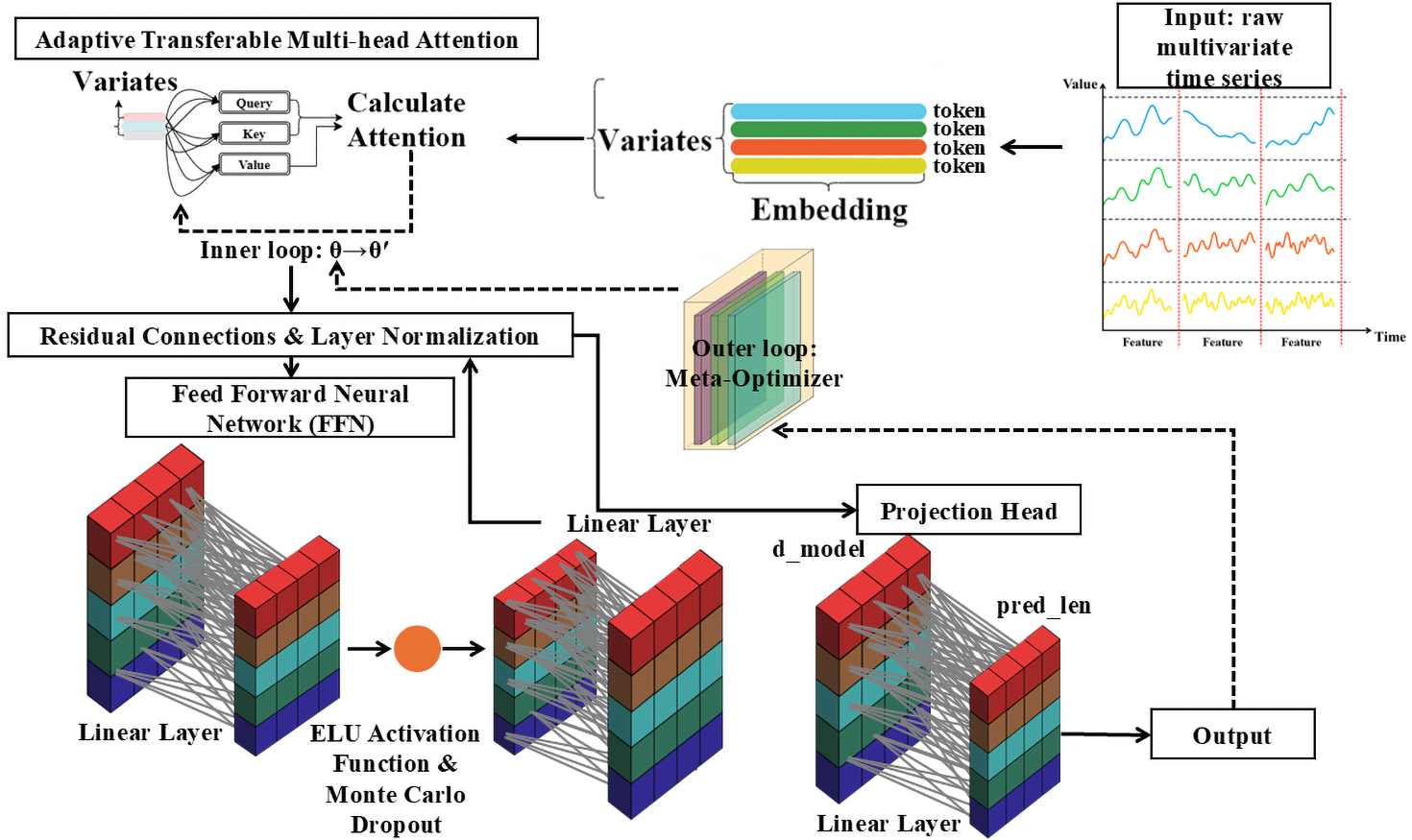}
    \caption{MMformer neural network flow chart, describing the core operating principle of the model and the operation mode of ATMA}
    \label{fig2}
\end{figure}

This subsection details our research methodology workflow. MMformer automates feature extraction from the dataset. During training, it employs meta-learning to learn sub-task-specific features. Subsequently, an outer loop updates shared parameters to identify and predict MTS feature patterns, as illustrated in Fig.\ref{fig2}.

The novelty of MMformer in workflow lies in the synergistic integration of key components, particularly in addressing the adaptability and generalization challenges inherent in complex multi-regional MTS prediction. Initially, MMformer processes the raw MTS data using a sliding-window approach to create input sequences and prediction targets, effectively generating subsets from the original data. The strategy enables the model to capture diverse temporal patterns, enhancing robustness and generalization capabilities.

A key architectural introduction involves the introduction of time stamps and dimension transposition, embedding each feature's time series into tokens. The design choice is intended to capture inter-variable correlations more effectively. The core of MMformer's innovation is the ATMA mechanism, which replaces the standard self-attention. ATMA uniquely integrates meta-learning (specifically, Model-Agnostic Meta-Learning, MAML). Through inner and outer optimization loops, ATMA effectively initializes and optimizes key parameters, substantially improving adaptability and generalization across various subtasks and data environments.

Furthermore, MMformer incorporates MC Dropout in the feed-forward networks (FFN), which enhances model generalization and robustness and provides uncertainty quantification, a capability not offered by traditional Dropout.

In summary, the innovation of MMformer resides in the elegant synergistic integration of deep learning (based on an encoder-only architecture) and meta-learning via the ATMA mechanism. This integration effectively addresses common challenges in complex MTS prediction, such as limited adaptability and the lack of uncertainty quantification. In contrast, it significantly reduces the reliance on manual feature engineering, improving precision and reliability.

\subsection{MMformer Enhancements and Innovations}
\subsubsection{Advancements in MMformer Architecture}
\begin{figure}[ht]
    \centering
    \includegraphics[width=1\textwidth]{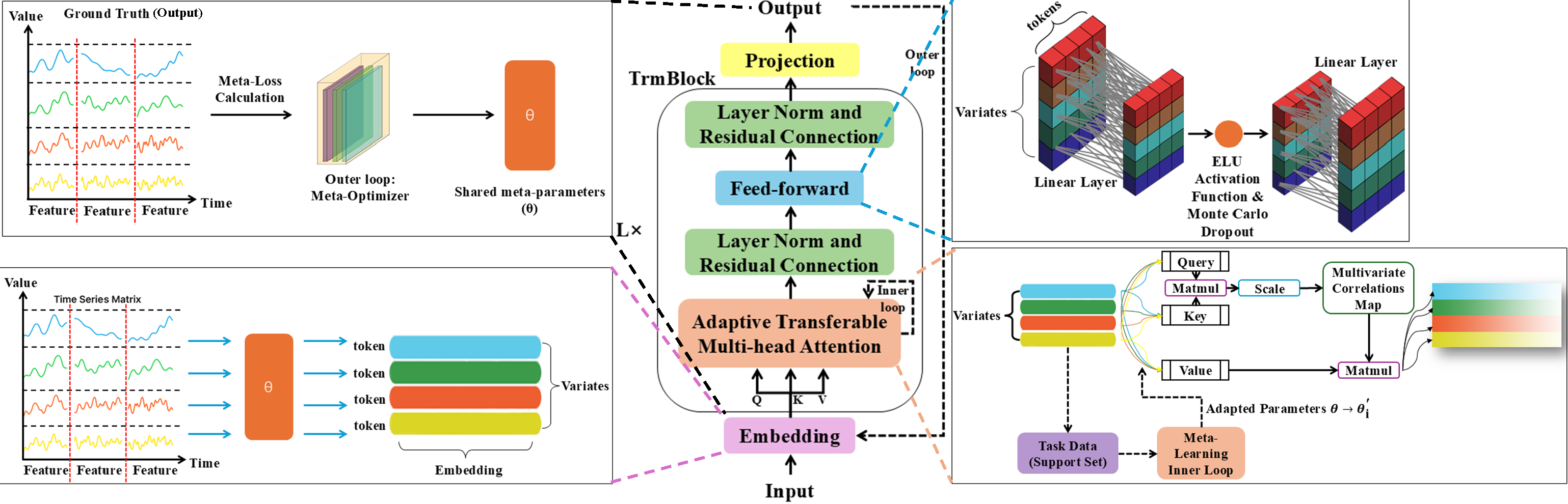}
    \caption{MMformer structure}
    \label{fig3}
\end{figure}

MMformer has the advantages of the iTransformer's architecture, enhances the self-attention mechanism, as illustrated in Fig.\ref{fig3}.
Compared with the traditional Transformer architecture and iTransformer incorporates a "Temporal LayerNorm" layer as defined in Eq.\ref{eq1} \citep{liu2023itransformer}:
\begin{equation}\label{eq1}
\text{LayerNorm}(H) = \left\{ \frac{h_n - \text{Mean}(h_n)}{\sqrt{\text{Var}(h_n)}} \mid n = 1, \dots, N \right\},
\end{equation}
where \( H \) is a matrix (\( N \)*\( D \)) of \( N \) tokens, each with dimension \( D \); \( h_n \) is the sequence vector for token \( n \), with \(\text{Mean}(h_n)\) and \(\text{Var}(h_n)\) representing its mean and variance, respectively. The purpose of expressing this operation in set form is to emphasize that standardization occurs separately for each time step (token). The set representation indicates that the same normalization procedure is applied independently to each time step $n$. 

Unlike conventional linear embedding approaches, MMFormer utilizes MLPs as the fundamental building blocks for both the embedding and mapping layers. The architecture utilizes within its embedding and mapping layers, processing input sequences into tokens that effectively encode both temporal and feature information. The methodology significantly improves the model's capacity to handle diverse input types and complex temporal dynamics with greater precision. MMformer enhances MTS analysis by integrating explicit time stamp embeddings and a refined temporal positional encoding scheme. The dual approach allows the model to capture not only the relative order of data points but also their absolute temporal context.

ATMA enables the model to learn how to learn by performing quick parameter updates on a support set within an inner training loop, dynamically synchronizing parameter adjustments throughout the model, significantly enhancing its predictive capacity and generalization ability on unseen tasks. ATMA allows MMformer to leverage information from the subtasks, improving its ability to generalize to new time series forecasting tasks and significantly enhancing predictive accuracy.

\begin{algorithm}[H]
\caption{MMformer Architecture: ATMA and MC Dropout}
\label{alg1}
\textbf{Require:} Input sequences $\mathbf{X} \in \mathbb{R}^{B \times L \times C}$; Model config \$C\$; Loss function $\mathcal{L}$; Learning rates $\alpha$ (inner), $\beta$ (outer). \\
\textbf{Ensure:} Predicted sequence $\mathbf{Y} \in \mathbb{R}^{B \times L' \times C}$.
\begin{algorithmic}[1]
\State Initialize model parameters $\Theta$ randomly.
\State \textbf{for} epoch = 1 to $N_{\text{epochs}}$ \textbf{do}
    \State \quad Randomly sample a mini-batch of tasks  $\mathcal{T}_i \sim p(\mathcal{T})$ from the training tasks.
    \State \quad \textbf{for} each task $\mathcal{T}_i$ in the mini-batch \textbf{do}
        \State \quad \quad \textbf{Inner Loop: Task Adaptation}
        \State \quad \quad Sample support set $\mathcal{D}_i^{support} = \{(\mathbf{x}_j, \mathbf{y}_j)\}_{j=1}^{K}$ and query set $\mathcal{D}_i^{query} = \{(\mathbf{x}_k, \mathbf{y}_k)\}_{k=1}^{K'}$ from $\mathcal{T}_i$.
        \State \quad \quad Initialize task-specific parameters $\Theta_i' \leftarrow \Theta$.
        \State \quad \quad \textbf{Feature Embedding:}
        \State \quad \quad \quad Permute $\mathbf{X}$ to $\mathbb{R}^{B \times C \times L}$ (features become tokens).
        \State \quad \quad \quad Embed temporal sequences within each feature; add positional encodings.
        \State \quad \quad \quad $E_{\text{enc}} \gets \text{DataEmbeddingInverted}(\mathbf{X}, C)$ \Comment{Embeds features to $d_{\text{model}}$; attention operates across features.}
        \State \quad \quad \textbf{Encoder Stack:} \textbf{for} $i = 1$ to $N_{inner\_steps}$ \textbf{do}
            \State \quad \quad \quad $E_{\text{enc}} \gets \text{EncoderLayer}_i(E_{\text{enc}})$
            \State \quad \quad \quad Integrates ATMA for adaptive, transferable attention: \Comment{Apply Eqs.\ref{eq5}-\ref{eq8}}
            \State \quad \quad \quad Compute  $\nabla_{\Theta_i'} \mathcal{L}_{\mathcal{T}_i}(\mathbf{X}, \mathbf{Y}; \Theta_i')$ using $\mathcal{D}_i^{support}$.
            \State \quad \quad \quad Update parameters: $\Theta_i' \leftarrow \Theta_i' - \alpha \nabla_{\Theta_i'} \mathcal{L}_{\mathcal{T}_i}$. \Comment{Apply Eq.\ref{eq9}}
        \State \quad \quad \textbf{End Inner Loop}
        \State \quad \quad \textbf{Prediction Projection:}
        \State \quad \quad \quad Project contextualized features ($d_{\text{model}}$) to \$L'\$ prediction length, then permute output.
        \State \quad \quad \quad Apply MC Dropout. \Comment{Apply Eqs.\ref{eq3}-\ref{eq4}}
        \State \quad \quad Compute meta-loss  $\mathcal{L}_{\text{meta}} = \mathcal{L}_{\mathcal{T}_i}(\mathbf{X}, \mathbf{Y}; \Theta_i')$ using $\mathcal{D}_i^{query}$  and adapted parameters $\Theta_i'$.
    \State \quad \textbf{end for}
    \State \quad \textbf{Meta-Objective:} Minimize meta-learning objective $\sum_{T_i \sim p(\mathcal{T})} \mathcal{L}_{T_i}\big(f_{\Theta_i'}\big)$. \Comment{Apply Eq.\ref{eq11}}
    \State \quad \textbf{Outer Loop: Meta-Optimization}
    \State \quad Update model parameters: $\Theta \leftarrow \Theta - \beta \nabla_{\Theta} \sum_{\mathcal{T}_i \sim p(\mathcal{T})} \mathcal{L}_{\text{meta}}$. \Comment{Apply Eq.\ref{eq10}}
\State \textbf{end for}
\State \textbf{Output:} Return $\mathbf{Y} = \text{Projector}(M)$ \Comment{Projects and permutes to prediction format.}
\end{algorithmic}
\end{algorithm}
To boost the model's robustness and representational power,  MC Dropout in the FFN replaces standard Dropout (see Eqs.\ref{eq3} and \ref{eq4}), which enables the capture of uncertainties in feature representations by performing multiple sampling during inference, which leads to more reliable predictions. During inference, multiple stochastic forward passes with dropout produce the average prediction \( \hat{y}\):
\begin{equation}\label{eq3}
   \centering
     \hat{y} = f(x; \theta, \text{dropout}),
\end{equation}
By performing multiple stochastic forward passes (e.g., $ T $ times) to obtain the average prediction $\bar{y}$:
\begin{equation}\label{eq4}
   \centering
     \bar{y} = \frac{1}{T} \sum_{t=1}^{T} \hat{y}^t,
\end{equation}
where $\hat{y}^t$ represents the output from the $ t $-th pass with dropout applied.

Algorithm \ref{alg1} presents the operational logic of advancements in MMformer architecture, detailing its innovative structure. The pseudocode outlines data embedding, sequence encoding, ATMA, and global MC Dropout. Algorithm \ref{alg1} also shows the time stamp with position and the Causal Mask mechanism of the architecture. MMformer's architecture, with its special embedding, ATMA, and MC Dropout, constitutes a powerful novel model for MTS analysis. The innovative approach enhances predictive accuracy and strengthens the model's generalization capabilities, offering a novel perspective and methodology for addressing complex time series problems.

\subsubsection{Adaptive transferable multi-head attention}
The self-attention mechanism encodes each input sequence element as a high-dimensional vector. For a sequence $X = (x_1, x_2, \dots, x_n)$, linear transformations generate query $Q$, key $K$, and value $V$ vectors as $Q = XW_Q$, $K = XW_K$, and $V = XW_V$ \citep{vaswani2023attention}.

The attention weight $\alpha^{(i,j)}$ between query $q_i$ at position $i$ and key $k_j$ at position $j$ is computed using the scaled dot-product:
\begin{equation} \label{eq5}
\alpha^{(i,j)} = \frac{q_i \cdot k_j}{\sqrt{d_k}},
\end{equation}
where $d_k$ is the dimension of $q_i$,. This scaling mitigates the softmax gradient issues for small values. The weights are then normalized via softmax:
\begin{equation} \label{eq6}
\alpha^{(i,j)} = \mathrm{softmax}(\alpha^{(i,j)}).
\end{equation}
The normalized weights $\alpha^{(i,j)}$ determine the contribution of each value vector $v_j$ to the representation $h_i$ at position $i$:
\begin{equation} \label{eq7}
h_i = \sum_j \alpha^{(i,j)}v_j.
\end{equation}
where Eq.\ref{eq7} multiplies the normalized attention weight $\alpha^{(i,j)}$ and the vector $v_j$ and sums each product up to obtain the weighted sum representation $h_i$ of the position $i$. It can be seen as a weighted sum of the vectors of all positions, and the weight is $\alpha^{(i,j)}$. 

Different from the standard Transformer, iTransformer's self-attention focuses on inter-variable rather than temporal correlations:
\begin{equation} \label{eq8}
\text{Attention}(Q, K, V) = \text{softmax}\left(\frac{QK^\top}{\sqrt{d_v}}\right)V,
\end{equation}
where $d_v$ is the dimensionality of the value vectors, while the mathematical formulations are similar, iTransformer emphasizes variable relationships over time steps, enhancing its capability to model complex multivariate dependencies.

\begin{figure}[ht]
    \centering
    \includegraphics[width=1\textwidth]{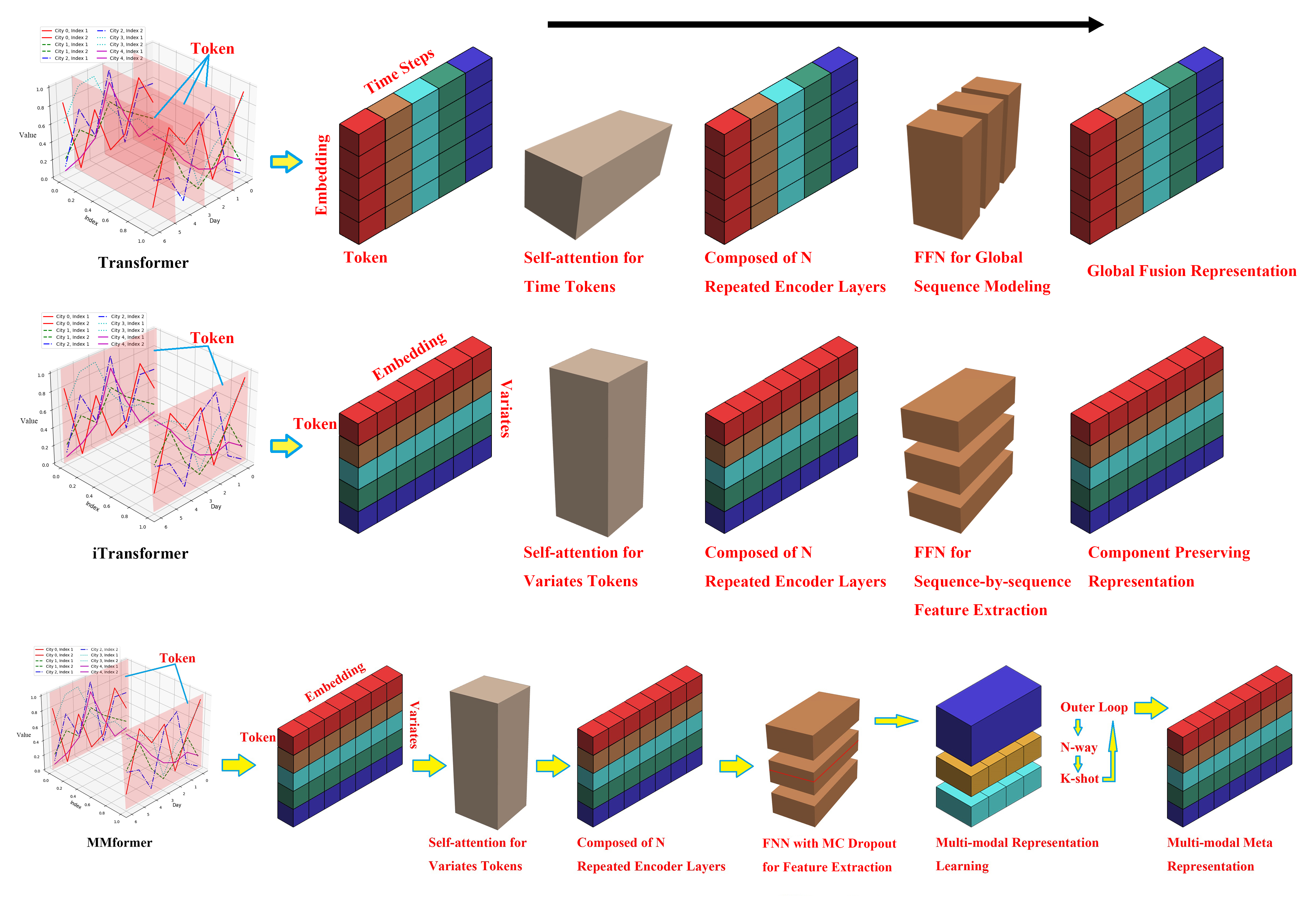}
    \caption{Comparison between Transformer, iTransformer, and MMformer architectures}
    \label{fig4}
\end{figure}

Fig.\ref{fig4} illustrates the distinct implementations of Transformer, iTransformer, and MMformer. Traditional Transformers aggregate multiple variables from a single time step into one token, capturing temporal dependencies but potentially hiding variable interactions. In contrast, iTransformer embeds each variable's entire MTS into separate tokens, allowing the attention mechanism to effectively capture inter-variable correlations. This approach facilitates non-linear representations through FFN, enhancing variable-centric learning and mitigating limitations associated with mixed time tags and limited receptive fields. However, the MMformer structure assigns independent tokens to each variable, which uses MLPs for embedding to capture the nonlinear characteristics of the data and incorporates the ATMA mechanism. The design improves MMformer's ability to process complex multi-regional MTS, maintaining robustness on small datasets or those with outliers, and only requires minimal parameter tuning.

The ATMA module defines a set of tasks $\mathcal{T}$, where each task $T_i \in \mathcal{T}$ includes a loss function $\mathcal{L}_{T_i}$, initial observation distribution $q(x_1)$, transition distribution $q(x_{t+1}|x_t, a_t)$, and designated time step $H_i$. A task $T_i$ is randomly selected, and model parameters $\theta$ are updated to $\theta_i'$ via gradient descent using loss function $\mathcal{L}_{T_i}$:
\begin{equation} \label{eq9}
\theta_i' = \theta - \alpha \nabla_{\theta} \mathcal{L}_{T_i}(f_{\theta}(x_{T_i}), y_{T_i}).
\end{equation}
where $\theta$ denotes the initial parameters for the current meta-iteration. $\theta_i'$ denotes the updated model parameters for task $T_i$ using task-specific data in the inner loop, and the adjustments at the task level are implicit, primarily focusing on updating the meta-parameters $\theta$ rather than directly computing or using each $\theta_i'$. $\alpha$ is the task-level learning rate that controls the model's learning step size on a single task. $\mathcal{L}_{T_i}$ is the loss function of task $T_i$ used to evaluate the $f_{\theta}$ performance on that task. $\nabla_{\theta} \mathcal{L}_{T_i}(f_{\theta}(x_{T_i}), y_{T_i})$ is the gradient of the loss function $\mathcal{L}_{T_i}$ to the model parameters $\theta$, where $x_{T_i}$ and $y_{T_i}$ represent the input data and target output of task $T_i$ respectively.

\begin{algorithm}[H]
\caption{Adaptive Transferable Multi-Head Attention (ATMA)}
\label{alg2}
\begin{algorithmic}[1]
\Require
    Query $\mathbf{Q}$, Key $\mathbf{K}$, Value $\mathbf{V}$, Attention mask $\mathbf{M}$,
    Number of heads \$H\$, Model dimension \$d\$, Meta-parameters $\theta$,
    Learning rate $\alpha$ (inner), $\beta$ (outer)
\Ensure
    Output context $\mathbf{O}$, Attention weights $\mathbf{A}$
\State \textbf{Compute per-head dimension:} $d_k = d/H$

\State \textbf{Step 1: Linear Projection}
    \Statex \hspace{1em} $\mathbf{Q}_h = \text{Linear}_Q(\mathbf{Q};\,\theta_Q)$, reshape to $(B, H, L_q, d_k)$
    \Statex \hspace{1em} $\mathbf{K}_h = \text{Linear}_K(\mathbf{K};\,\theta_K)$, reshape to $(B, H, L_k, d_k)$
    \Statex \hspace{1em} $\mathbf{V}_h = \text{Linear}_V(\mathbf{V};\,\theta_V)$, reshape to $(B, H, L_k, d_k)$

\State \textbf{Step 2: Meta-Learning Inner Loop (\textit{Cross-task Adaptation})}
    \For{each task $T_i$ in current meta-batch}
        \State Compute inner loss: $\mathcal{L}_{T_i}^{\text{train}} = \mathcal{L}(\text{ATMA}(\mathbf{Q},\mathbf{K},\mathbf{V};\theta),~\text{label}_{T_i})$
        \State \textit{(Meta-learning adaptation)} Update task-specific parameters:
        \Statex \hspace{2em} $\theta'_{T_i} \leftarrow \theta - \alpha \nabla_\theta \mathcal{L}_{T_i}^{\text{train}}$ \Comment{Apply Eq.\ref{eq9}}
    \EndFor

\State \textbf{Step 3: Scaled Dot-Product Attention with MC Dropout}
    \State For each task $T_i$:
    \Statex \hspace{1em} $S = \mathbf{Q}_h \cdot \mathbf{K}_h^\top / \sqrt{d_k}$  \Comment{Implicitly uses Eq.\ref{eq5}}
    \If{$\mathbf{M}$ is not None}
        \State Mask $S$ at the specified positions by $-\infty$
    \EndIf
    \State $\mathbf{A} = \text{Softmax}(S, \text{dim}=-1)$ \Comment{Implicitly uses Eq.\ref{eq6}}
    \If{training phase}
        \State Apply MC Dropout: $\mathbf{A} \leftarrow \text{MC Dropout}(\mathbf{A})$ \Comment{Apply Eqs.\ref{eq3}-\ref{eq4}}
    \EndIf

\State \textbf{Step 4: Multi-head Context Composition and Output}
    \Statex \hspace{1em} $\mathbf{C}_h = \mathbf{A} \cdot \mathbf{V}_h$  \Comment{Applies Eq.\ref{eq7}}
    \Statex \hspace{1em} Concatenate $\mathbf{C}_h$ across all heads, reshape to $(B, L_q, d)$
    \Statex \hspace{1em} Output linear layer: $\mathbf{O} = \text{Linear}_O(\mathbf{C})$

\State \textbf{Step 5: Meta-Learning Outer Loop (\textit{Cross-task Parameter Update})}
    \For{each task $T_i$ in meta-batch}
        \State Compute validation loss: $\mathcal{L}_{T_i}^{\text{val}} = \mathcal{L}(\text{ATMA}(\mathbf{Q},\mathbf{K},\mathbf{V};\theta'_{T_i}),~\text{label}_{T_i}^{\text{val}})$
    \EndFor
    \State Aggregate meta-loss: $\mathcal{L}_{\text{meta}} = \frac{1}{|\mathcal{T}|} \sum_{T_i} \mathcal{L}_{T_i}^{\text{val}}$.
    \State Define Meta-Objective: the outer loop minimizes the meta-learning objective $\sum_{T_i \sim p(\mathcal{T})} \mathcal{L}_{T_i}\big(f_{\Theta_i'}\big)$. \Comment{Apply Eq.\ref{eq11}}
    \State Update shared parameters: $\theta \leftarrow \theta - \beta \nabla_\theta \mathcal{L}_{\text{meta}}$. \Comment{Apply Eq.\ref{eq10}}
\State \Return Output $\mathbf{O}$, Attention weights $\mathbf{A}$.
\end{algorithmic}
\end{algorithm}
The aim of this step is to optimize the parameter $\theta$ such that the sum of the loss functions across all tasks in the task set $\mathcal{T}$ is minimized, i.e., $\min_{\theta} \sum_{T_i \sim \mathcal{T}} \mathcal{L}_{T_i}(f_{\theta}(x_{T_i}), y_{T_i})$, where the notation $\sim$ indicates that tasks $T_i$ are sampled from the task set $\mathcal{T}$. The objective is to minimize the cumulative loss across all tasks:
\begin{equation} \label{eq10}
\theta \gets \theta - \beta \nabla_{\theta} \sum_{T_i \sim \mathcal{T}} \mathcal{L}_{T_i}(f_{\theta}(x_{T_i}), y_{T_i}),
\end{equation}
where $\beta$ is the meta-learning rate, which controls the learning step length of the model on multiple tasks. Eq.\ref{eq10} leverages the gradients obtained from all sampled tasks $T_i$ during the inner loop (Eq.\ref{eq9}) to adjust the meta-parameters $\theta$. In the outer loop, the overall model parameters $\theta$ are updated by aggregating the meta-gradients from multiple tasks, each adjusted by a specific learning rate $\beta$. Consequently, there is no single formula for $\theta'$, as it represents an intermediate result for each task during the inner loop execution.

The meta-learning objective function is:
\begin{equation} \label{eq11}
\sum_{T_i \sim \mathcal{T}} \mathcal{L}_{T_i}\big(f_{\theta - \alpha \nabla_{\theta} \mathcal{L}_{T_i}(f_{\theta}(x_{T_i}), y_{T_i})}\big),
\end{equation}
where aiming to find optimal $\theta$ that minimizes the cumulative loss through iterative inner loop (Eq.~\ref{eq9}) and outer loop (Eq.~\ref{eq10}) updates that reduce the cumulative loss across all tasks in the task set $\mathcal{T}$.

Algorithm \ref{alg2} details the logic of the ATMA mechanism. It specifies the input parameters and outputs, and systematically outlines the process, including Linear Projection, the Meta-Learning Inner Loop, Attention with MC Dropout, Multi-head Context Composition, and the Meta-Learning Outer Loop. Algorithm \ref{alg2} highlights the core innovation and comprehensively describes the entire process.

The proposed ATMA mechanism combines Probattention with meta-learning to enhance MTS analysis. By defining a diverse task set and performing gradient updates per task, ATMA enables rapid model adaptation and improves generalization across tasks through meta-gradients. This integration extends self-attention to multi-dimensional long and short-term sequences, resulting in the MMformer model. MMformer has robustness on limited or noisy datasets while remaining user-friendly for researchers lacking tuning experience. ATMA's innovative design offers significant potential for advancements in related domains.

\section{Experiments Application and Results}
This section employs the MMformer to analyze two real-world datasets and general MTS datasets, PEMS. The real-world datasets, sourced from the China Meteorological Data Service Center under the National Meteorological Information Center \citep{cmdc}, encompass comprehensive raw data. We benchmark MMformer against iTransformer, PatchTST, TimesNet, and Transformer to demonstrate its superior performance. The evaluation provides insights into MMformer's effectiveness and generalization as the MTS prediction tool for environmental monitoring.

\subsection{Multivariate Time Series Data Display}
The proposed MMformer model's effectiveness in predicting environmental change using daily average measurements of seven indicators across 331 Chinese cities over 1,277 days (January 2018 to June 2021) and temperature precipitation at 909 sites from 2018 to 2022 (1826 days). This research uses daily average values for all indicators rather than instantaneous measurements, which reduces the impact of transient phenomena on prediction outcomes. While environmental data contains ephemeral fluctuations, such as short-term pollution peaks or rapid meteorological changes, these phenomena are effectively smoothed in daily average aggregations, reducing the influence of extreme values on overall trends.

\begin{figure}[ht]
    \centering
    \includegraphics[width=0.9\textwidth]{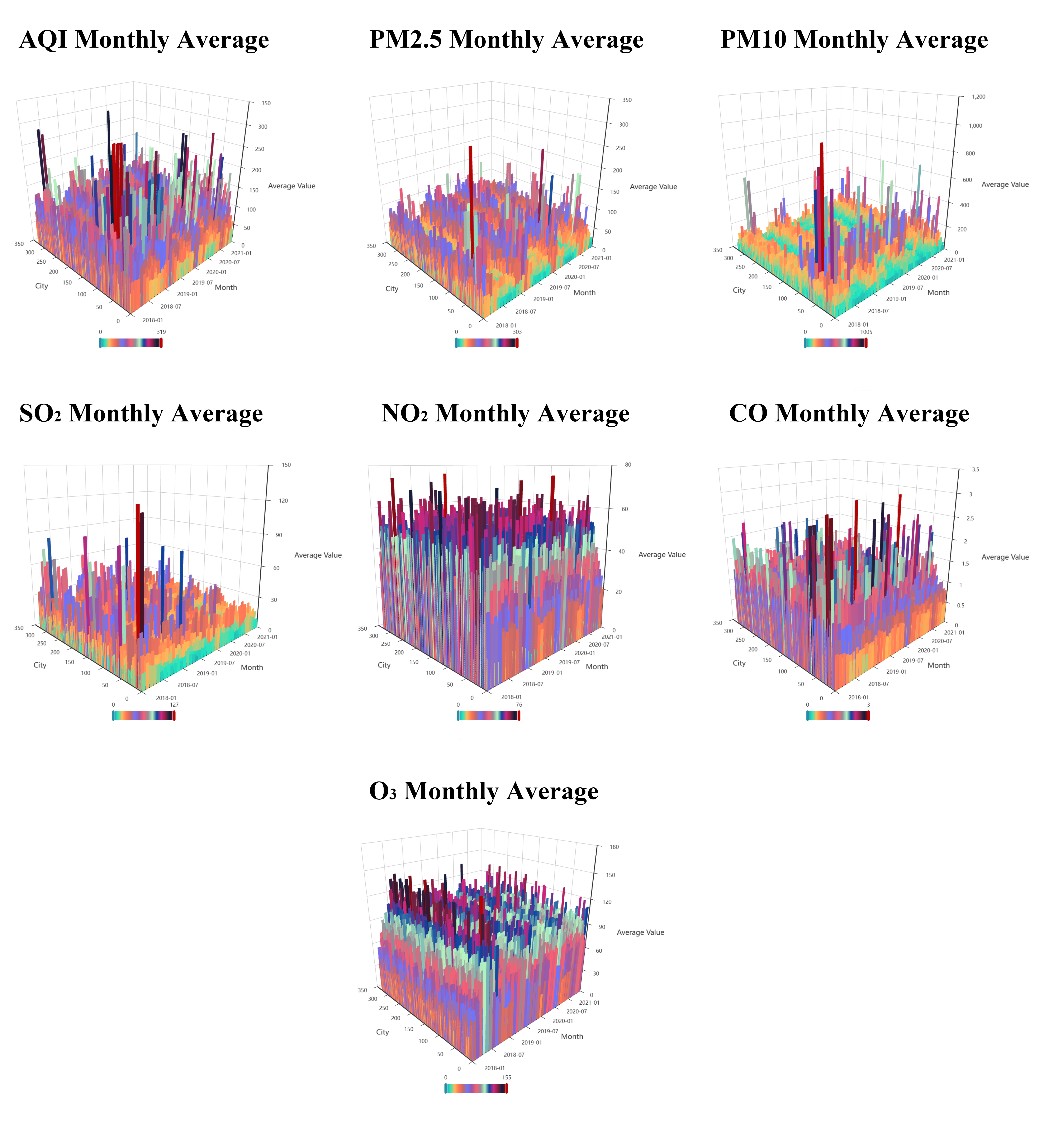}\\
    \caption{Air Quality Data from January 2018 to Jun 2021}
    \label{fig5}
\end{figure}

Fig.\ref{fig5} depicts the monthly air quality average values of each indicator in all cities to illustrate the general distribution and outliers of the data, providing a general understanding. The horizontal and vertical axes represent the city number, the corresponding month, and the monthly average value of the city at that time.

Table \ref{table1} summarizes the seven air quality indicators. After initial processing, the dataset's features are detailed in Table \ref{table2}, laying the foundation for subsequent analysis, which combines with Fig.\ref{fig9} we can observe the value ranges of the seven features vary greatly, and the mean, median, and 25\% quartile do not fluctuate much within four years, while the maximum values of some indicators change significantly.

\begin{table}[H]
    \centering
    \caption{Summary of features from meteorological MTS data of weather stations}
    \label{table1}
    \small
    \begin{tabularx}{\textwidth}{@{}>{\centering\arraybackslash}X>{\centering\arraybackslash}X@{}}
        \toprule
        \textbf{Feature} & \textbf{Description} \\
        \midrule
        AQI & Air quality index \\
        PM2.5 & Fine particulate matter less than or equal to 2.5 microns in diameter \\
        PM10 & Inhalable particulate matter with a diameter less than or equal to 10 microns \\
        $\mathrm{SO}_2$ & Sulfur dioxide can cause respiratory \\
        $\mathrm{NO}_2$ & Nitrogen dioxide can cause respiratory tract irritation \\
        $\mathrm{CO}$ & Carbon monoxide impair oxygen transport \\
        $\mathrm{O}_3$ & Ozone will irritate the respiratory \\
        \bottomrule
    \end{tabularx}
\end{table}

\begin{longtable}{@{\extracolsep{\fill}}lp{2cm}p{2cm}p{1.5cm}p{2cm}@{}}
    \caption{Summary of Air Quality Statistics about 7 indexes}
    \label{table2} \\
    \toprule
    \textbf{Indicator} & \textbf{Maximum Value} & \textbf{25\% QUARTILE} & \textbf{Mean} & \textbf{Median}\tnote{1} \\
    \midrule
    \endfirsthead
    \multicolumn{5}{c}%
    {\tablename\ \thetable\ -- \textit{Summary of Air Quality Statistics about 7 indexes}} \\
    \toprule
    \textbf{Indicator} & \textbf{Maximum Value} & \textbf{25\% QUARTILE} & \textbf{Mean} & \textbf{Median}\tnote{1} \\
    \midrule
    \endhead
    \midrule
    \multicolumn{5}{r}{\textit{Continued on next page}} \\
    \endfoot
    \bottomrule
    \endlastfoot
    2018 AQI & 500 & 45 & 71 & 61 \\
    2019 AQI & 500 & 43 & 67 & 58 \\
    2020 AQI & 500 & 33 & 57 & 48 \\
    2021 AQI & 500 & 37 & 65 & 54 \\
    2018 PM2.5 & 1787 & 19 & 39 & 30 \\
    2019 PM2.5 & 1348 & 19 & 38 & 30 \\
    2020 PM2.5 & 1188 & 15 & 33 & 25 \\
    2021 PM2.5 & 1399 & 17 & 36 & 28 \\
    2018 PM10 & 5850 & 38 & 76 & 58 \\
    2019 PM10 & 4709 & 38 & 70 & 58 \\
    2020 PM10 & 2767 & 30 & 60 & 48 \\
    2021 PM10 & 6593 & 34 & 77 & 55 \\
    2018 SO$_2$ & 251 & 6 & 13 & 10 \\
    2019 SO$_2$ & 241 & 6 & 11 & 9 \\
    2020 SO$_2$ & 142 & 5 & 10 & 8 \\
    2021 SO$_2$ & 432 & 5 & 10 & 8 \\
    2018 NO$_2$ & 163 & 16 & 27 & 24 \\
    2019 NO$_2$ & 131 & 16 & 28 & 25 \\
    2020 NO$_2$ & 125 & 14 & 25 & 21 \\
    2021 NO$_2$ & 152 & 14 & 24 & 22 \\
    2018 CO & 6.88 & 0.59 & 0.85 & 0.76 \\
    2019 CO & 5.49 & 0.57 & 0.80 & 0.73 \\
    2020 CO & 26.42 & 0.51 & 0.73 & 0.66 \\
    2021 CO & 13.76 & 0.50 & 0.71 & 0.65 \\
    2018 O$_3$ & 245 & 42 & 65 & 62 \\
    2019 O$_3$ & 276 & 40 & 61 & 58 \\
    2020 O$_3$ & 218 & 41 & 62 & 60 \\
    2021 O$_3$ & 216 & 45 & 64 & 62 \\
    2018-2021 AQI & 500 & 40 & 65 & 56 \\
    2018-2021 PM2.5 & 1787 & 17 & 37 & 28 \\
    2018-2021 PM10 & 6593 & 35 & 70 & 54 \\
    2018-2021 SO$_2$ & 432 & 6 & 11 & 8 \\
    2018-2021 NO$_2$ & 163 & 15 & 26 & 23 \\
    2018-2021 CO & 26.42 & 0.55 & 0.78 & 0.71 \\
    2018-2021 O$_3$ & 276 & 42 & 63 & 60 \\
\end{longtable}

\begin{table}[h]
    \centering
    \caption{Summary of Weather Station Statistics about Precipitation and Temperature}
    \label{table6}
    \begin{threeparttable}
        \begin{tabularx}{\linewidth}{@{}>{\bfseries}lcccc@{}}
            \toprule
            \textbf{Indicator} & \textbf{Maximum} & \textbf{Minimum} & \textbf{Mean} & \textbf{NSBM}\tnote{1} \\
            \midrule
            Total Precipitation\tnote{2} & 297.10 & 0.00 & 1.74 & 521 \\
            2018 Precip & 199.88 & 0.00 & 1.76 & 544 \\
            2019 Precip & 297.10 & 0.00 & 1.69 & 555 \\
            2020 Precip & 205.15 & 0.00 & 1.82 & 547 \\
            2021 Precip & 272.75 & 0.00 & 1.78 & 513 \\
            2022 Precip & 163.34 & 0.00 & 1.64 & 534 \\
            Total Temperature\tnote{3} & 37.08 & -40.80 & 7.17 & 463 \\
            2018 Temp & 35.56 & -40.80 & 6.96 & 467 \\
            2019 Temp & 36.01 & -38.24 & 7.19 & 460 \\
            2020 Temp & 34.08 & -38.40 & 7.06 & 465 \\
            2021 Temp & 36.32 & -40.58 & 7.32 & 463 \\
            2022 Temp & 37.08 & -37.57 & 7.34 & 465 \\
            \bottomrule
        \end{tabularx}
        \begin{tablenotes}
            \footnotesize
            \item[1] Number of sites below the mean
            \item[2] The unit of precipitation: 0.1mm
            \item[3] The unit of temperature: 1℃
        \end{tablenotes}
    \end{threeparttable}
\end{table}
Table \ref{table6} provides a statistical summary of the processed dataset, encompassing precipitation and temperature metrics within the climate dataset. Comprehensive analysis of Table \ref{table6} reveals highly heterogeneous rainfall distribution across the five years, contrasted with consistently rising temperatures.

\subsection{Evaluation Baseline Metrics}
Appropriate evaluation metrics are essential to assess the performance of various models in MTS prediction tasks. This study employs Mean Absolute Error (MAE), Mean Squared Error (MSE), and Mean Absolute Percentage Error (MAPE) as the primary indicators, each providing distinct insights into prediction accuracy and deviation.

\begin{itemize}
    \item[$\bullet$]
    \textbf{Mean Absolute Error (MAE)} \citep{KARUNASINGHA2022609}: MAE calculates the average absolute differences between predicted and actual values, offering a straightforward measure of prediction accuracy. A lower MAE indicates better model performance.
    \begin{equation} \label{eq12}
    \text{MAE} = \frac{1}{n} \sum_{i=1}^{n} |y_i - \hat{y}_i|
    \end{equation}
    where $y_i$ are the actual values, $\hat{y}_i$ the predicted values, and $n$ the number of observations.

    \item[$\bullet$]
    \textbf{Mean Squared Error (MSE)} \citep{KARUNASINGHA2022609}: MSE measures the average of the squared differences between predicted and actual values, making it sensitive to larger errors. The sensitivity helps identify models with lower variance in their predictions.
    \begin{equation} \label{eq13}
    \text{MSE} = \frac{1}{n} \sum_{i=1}^{n} (y_i - \hat{y}_i)^2
    \end{equation}
Each error $(y_i - \hat{y}_i)$ is squared in this formula, amplifying more significant errors. It means the MSE heavily penalizes significant deviations from the actual values, highlighting models with lower prediction variance.

    \item[$\bullet$]
    \textbf{Mean Absolute Percentage Error (MAPE)} \citep{efendi2023role}: MAPE expresses the accuracy as a percentage, facilitating easy comparison across different datasets. It calculates the average absolute percentage deviations between predicted and actual values, with lower values indicating higher accuracy.
    \begin{equation} \label{eq14}
    \text{MAPE} = \frac{1}{n} \sum_{i=1}^{n} \left| \frac{y_i - \hat{y}_i}{y_i} \right| \times 100\%
    \end{equation}
\end{itemize}

\subsection{Results Analysis on Datasets}
The predictive analysis of data consists mainly of two stages: fundamental statistical analysis and comparison of the prediction performance of various methods.

\subsubsection{Basic Statistical Analysis of Data}
Table \ref{table3} and Table \ref{table7} summarise the statistics of key indicators in the air quality and climate datasets. Based on Table \ref{table3} analysis, all air quality features show minimum values of 0, with significant variance across all indicators except $CO$, indicating substantial regional air quality disparities. The consistently lower median values compared to means suggest right-skewed distributions, where generally good air quality is punctuated by severe pollution events. These spatial and temporal variations in pollution levels underscore the necessity for robust, targeted monitoring systems in high-risk areas.
\begin{table}[H]
    \centering
    \caption{Basic statistical analysis of air quality characteristic indicators}
    \label{table3}
    \begin{threeparttable}
        \small
        \begin{tabularx}{\textwidth}{@{}>{\bfseries}lXXXXXXX@{}}
            \toprule
            \textbf{Indicator} & \textbf{Mean} & \textbf{Variance} & \textbf{Max} & \textbf{Min} & \textbf{Median}\\
            \midrule
            AQI & 65 & 1842 & 500 & 0 & 56 \\
            PM2.5 & 37 & 1076 & 1787 & 0 & 28 \\
            PM10 & 70 & 5797 & 6593 & 0 & 54 \\
            SO$_2$ & 11 & 88 & 432 & 0 & 8 \\
            NO$_2$ & 26 & 231 & 163 & 0 & 23 \\
            CO & 0.78 & 0.14 & 26.42 & 0 & 0.71 \\
            O$_3$ & 63 & 847 & 276 & 0 & 60 \\
            \bottomrule
        \end{tabularx}
    \end{threeparttable}
\end{table}

Table \ref{table7} presents a statistical overview of precipitation and temperature data from 2018 to 2022. Precipitation means are comparable across these years, although annual variances and interquartile ranges (IQRs) show differences, with 2020 exhibiting the highest variance and 2018 the greatest IQR for precipitation. Temperature means are also consistent throughout the study period. However, temperature displayed more significant variability in 2018, as indicated by the highest variance and IQR, suggesting greater dispersion in middle temperature values. Overall, the analysis highlights a degree of interannual stability in these climate indicators.
\begin{table}[H]
    \centering
    \caption{Basic statistical analysis of temperature and precipitation characteristic indicators}
    \label{table7}
    \begin{threeparttable}
        \small
        \begin{tabularx}{\textwidth}{@{}>{\bfseries}lXXXXXXX@{}}
            \toprule
            \textbf{Indicator} & \textbf{Mean} & \textbf{Variance} & \textbf{Mode} & \textbf{Range} & \textbf{IQR}\tnote{1} \\
            \midrule
            Precipitation\tnote{2} & 1.74 & 28.90 & 0.00 & 297.10 & 0.99\\
            2018\tnote{3}  & 1.76 & 27.21 & 0.00 & 199.88 & 1.06 \\
            2019  & 1.69 & 28.23 & 0.00 & 297.10 & 0.97 \\
            2020  & 1.82 & 33.39 & 0.00 & 205.15 & 0.96 \\
            2021  & 1.78 & 29.89 & 0.00 & 272.75 & 1.03 \\
            2022  & 1.64 & 25.74 & 0.00 & 163.34 & 0.95 \\
            Temperature\tnote{4} & 7.17 & 173.32 & 11.49 & 77.88 & 20.24 \\
            2018  & 6.96 & 181.39 & 5.38 & 76.35 & 20.71 \\
            2019  & 7.19 & 168.18 & -17.77 & 74.25 & 20.20 \\
            2020  & 7.06 & 167.48 & 6.56 & 72.48 & 20.14 \\
            2021  & 7.32 & 170.14 & 9.24 & 76.90 & 19.79 \\
            2022  & 7.34 & 179.30 & 6.19 & 74.66 & 20.39 \\
            \bottomrule
        \end{tabularx}
        \begin{tablenotes}
            \footnotesize
            \item[1] Interquartile range
            \item[2] The unit of precipitation: 0.1mm
            \item[3] The unit of time: year
            \item[4] The unit of Temperature: 1℃
        \end{tablenotes}
    \end{threeparttable}
\end{table}

\subsubsection{Comparative Analysis of Prediction Methods on Real-World Data}
\paragraph{Comparative Analysis on the Air Quality Dataset}
We compare the proposed MMformer with four recent strong baseline models, iTransformer, PatchTST, TimesNet, and Transformer, using MSE, MAE, and MAPE as primary evaluation metrics. The reason for selecting these baseline models is that the ICLR 2024 Spotlight SOTA model, iTransformer, refines the Transformer architecture to enhance robustness and accuracy in MTS and achieves state-of-the-art performance in general MTS datasets \citep{liu2023itransformer}. PatchTST is recognized as a strong baseline for its innovative patching strategy, which divides time series into segments and applies a Transformer to capture both local dependencies within patches and global context across the series, demonstrating strong performance on diverse MTS benchmarks \citep{zhang2024patchtcn}. Similarly, TimesNet is highlighted for its unique approach to time series decomposition, effectively separating trend and seasonal components to better model complex temporal patterns, thereby achieving leading results on various benchmark tasks \citep{wu2023timesnettemporal2dvariationmodeling}. Because of its widely used self-attention mechanism, Transformer is considered one of the benchmark models in MTS forecasting \citep{mendis2024multivariate}. Including these strong baseline models highlights the MMformer in environmental multi-regional MTS prediction.
\begin{table}[H]
    \centering
    \caption{Performance Comparison of Multivariate Time Series Forecasting Models on the Air Quality Dataset. The seq\_len and pred\_len are set to 150 and 30 for all models. The results are rounded to three decimal places. Figures in red and bold are the best for that metric.}
    \label{table4}
    \begin{threeparttable}
        \small
        \begin{tabularx}{\textwidth}{@{}>{\bfseries}lXXXXXXX@{}}
            \toprule
            \textbf{Model} & \textbf{MSE} & \textbf{MAE} & \textbf{MAPE}\\
            \midrule
            MMformer & \textcolor{red}{\textbf{0.245}} & \textcolor{red}{\textbf{0.332}} & \textcolor{red}{\textbf{18.872}}\%\\
            iTransformer & 0.822 & 0.529 & 29.738\%\\
            TimesNet & 0.807 & 0.528 & 29.319\%\\
            PatchTST & 0.862 & 0.556 & 31.133\%\\
            Transformer & 0.770 & 0.524 & 29.188\%\\
            \bottomrule
        \end{tabularx}
    \end{threeparttable}
\end{table}
Table \ref{table4} presents the results of MTS prediction on a real air quality dataset, evaluated using MSE, MAE, and MAPE metrics. Experimental results demonstrate that MMformer achieves the lowest MSE (0.245), MAE (0.332), and MAPE (18.872\%), indicating a superior ability for error minimization and data fitting. A detailed comparative analysis reveals substantial improvements over all baseline methods. Compared to iTransformer, MMformer reduces MSE by 70.195\%, MAE by 37.240\%, and MAPE by 36.542\%. Regarding TimesNet, MMformer decreases MSE by 69.641\%, MAE by 37.121\%, and MAPE by 35.632\%. Compared to PatchTST, MMformer reduces MSE by 71.578\%, MAE by 40.288\%, and MAPE by 39.383\%. Finally, relative to the Transformer model, the proposed MMformer shows significant reductions in MSE (decreased 68.182\%), MAE (decreased 36.641\%), and MAPE (decreased 35.343\%). These findings strongly emphasize the MMformer's robustness and effectiveness in predicting complex, high-dimensional air quality data, thus showcasing its substantial potential for practical environmental monitoring and forecasting applications.

The MMformer model incorporates MC Dropout for uncertainty quantification, thereby enabling the evaluation of prediction reliability. On the air quality dataset, MMformer exhibits a global average variance of 0.011919 and a global average standard deviation of 0.105612, reflecting the overall dispersion of its predictions. Across all samples, time steps, and features, the variance ranges from a minimum of 0.005076 to a maximum of 0.146630. The average 95\% confidence interval width, estimated using the global average standard deviation, is approximately 0.4142 ($2 \times 1.96 \times 0.105612$). While the low global average variance and interval width suggest robust overall prediction confidence, the identified maximum variance of 0.146630 clearly indicates specific scenarios with reduced model confidence. Crucially for risk assessment, the maximum variance value directly flags the most unreliable prediction points as high-risk regions. In risk-sensitive applications, such as energy load forecasting or environmental disaster prediction, identifying these high-variance areas is paramount for effective risk management and optimized decision-making. Thus, MMformer's variance analysis serves as a vital tool for reliability assessment and supports informed risk management strategies.

\begin{figure}[ht]
    \centering
    \includegraphics[width=0.9\textwidth]{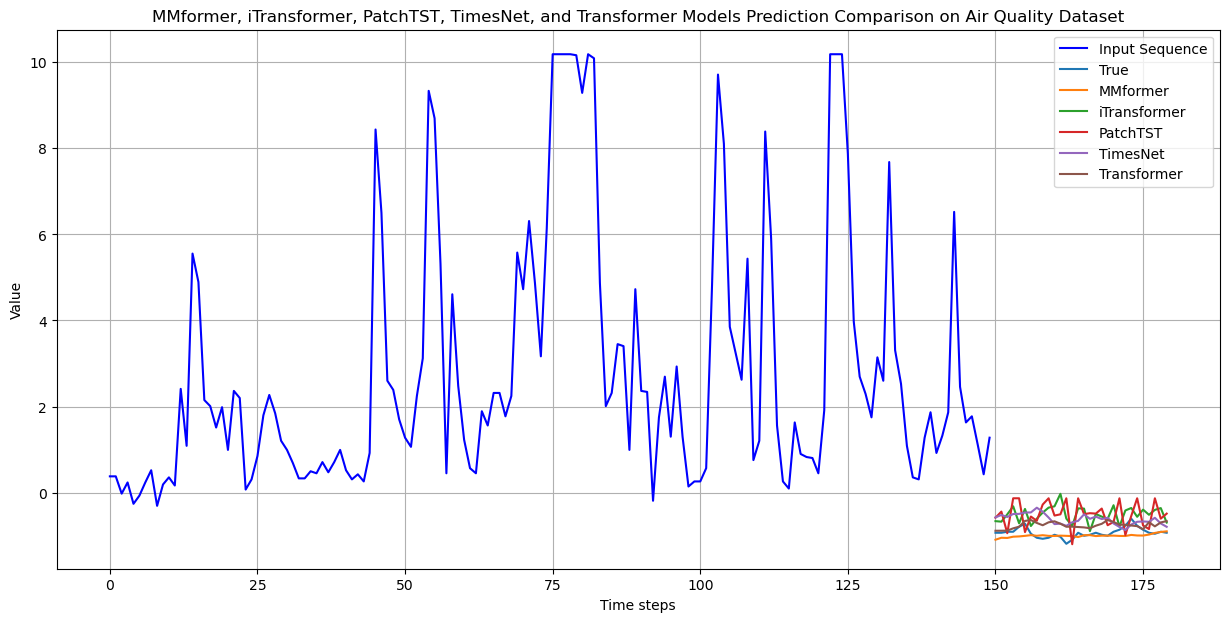}\\
    \caption{MMformer, iTransformer, PatchTST, TimesNet, and Transformer Models Prediction Comparison on Air Quality Dataset}
    \label{fig6}
\end{figure}
For data preparation, we allocated 730 days for training, 366 days for validation, and 181 days for testing, employing a sliding window strategy to capture temporal dependencies. Sequence lengths = 150, prediction lengths = 30, and hyperparameters are carefully aligned across all models to ensure a fair comparison. Fig.\ref{fig6} demonstrates MMformer's overall advantage in capturing intrinsic properties of MTS data by the first feature sample of all baseline models. The visualization reveals that MMformer (orange line) most accurately tracks the true values (light blue line) throughout the prediction horizon, effectively capturing both gradual trends and sudden fluctuations. In contrast, other baseline models exhibit larger deviations from actual observations, particularly during periods of rapid change, highlighting MMformer's superior capability in modeling complex climate patterns.

\begin{figure}[H]
    \centering
    \includegraphics[width=1\textwidth]{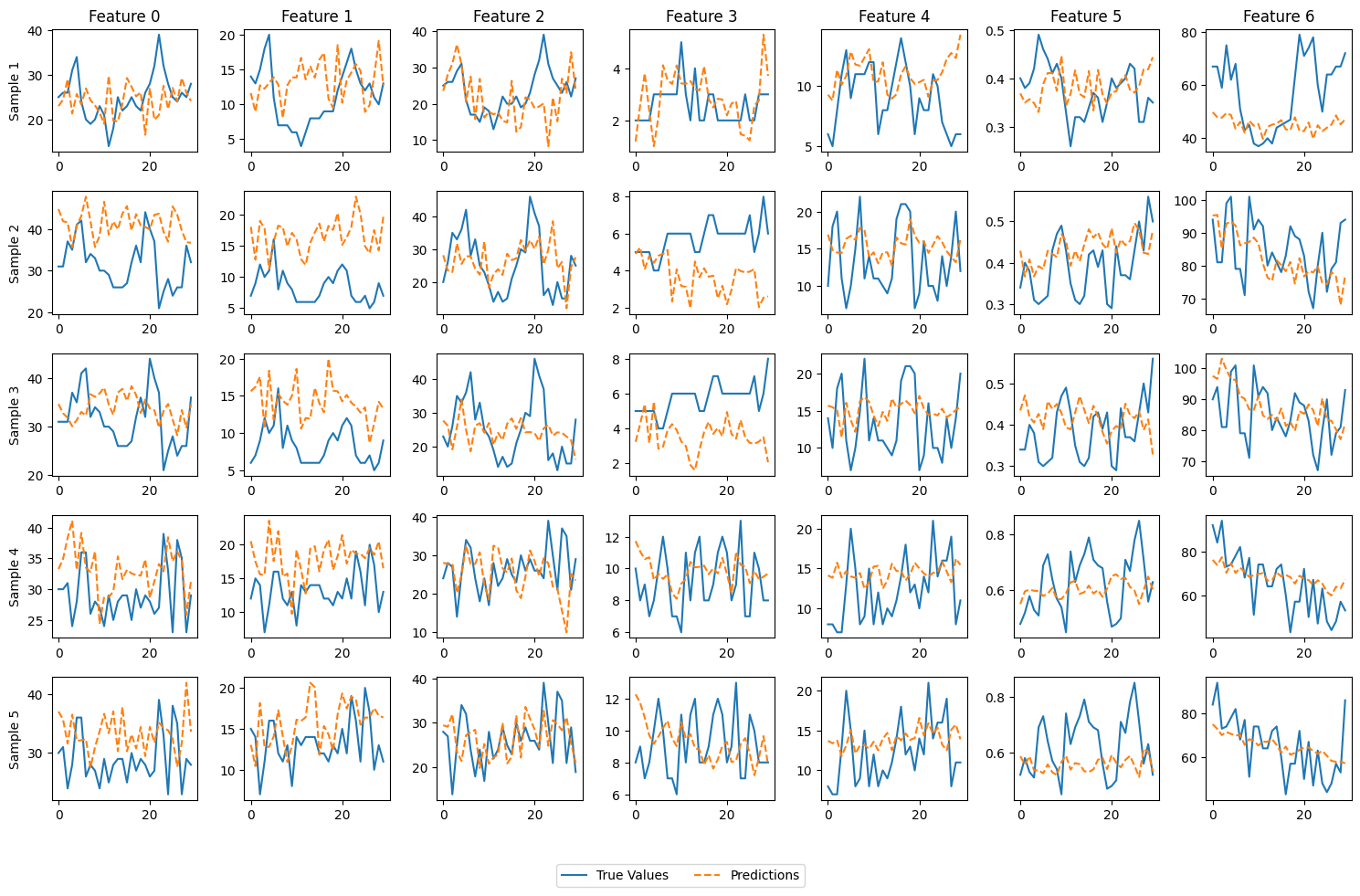}\\
    \caption{MMformer prediction in the test dataset with the five test samples with the lowest MSE in the Air Quality Dataset}
    \label{fig7}
\end{figure}

\begin{figure}[H]
    \centering
    \includegraphics[width=1\textwidth]{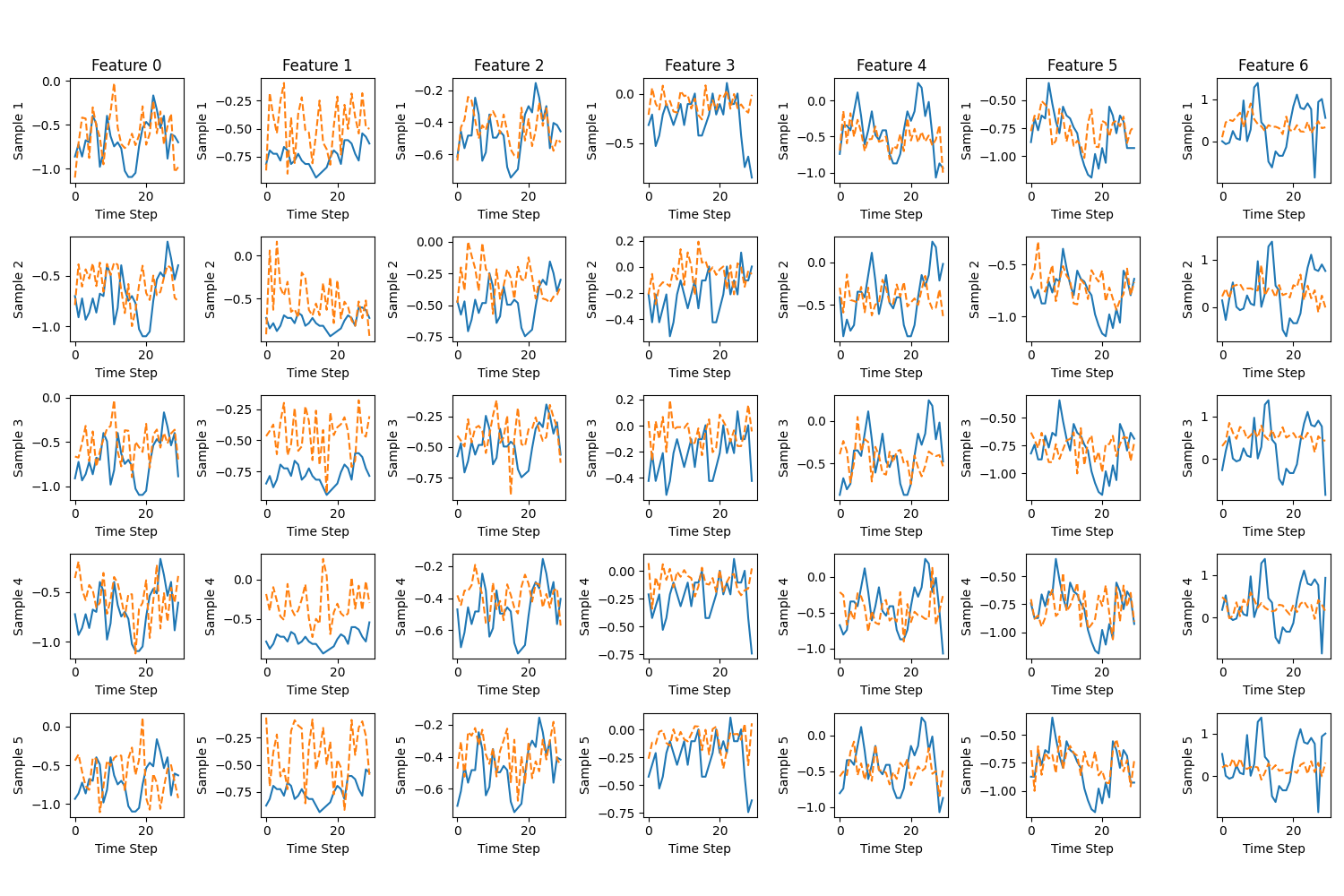}\\
    \caption{iTransformer prediction in the test dataset with the five test samples with the lowest MSE in the Air Quality Dataset}
    \label{fig8}
\end{figure}

\begin{figure}[H]
    \centering
    \includegraphics[width=1\textwidth]{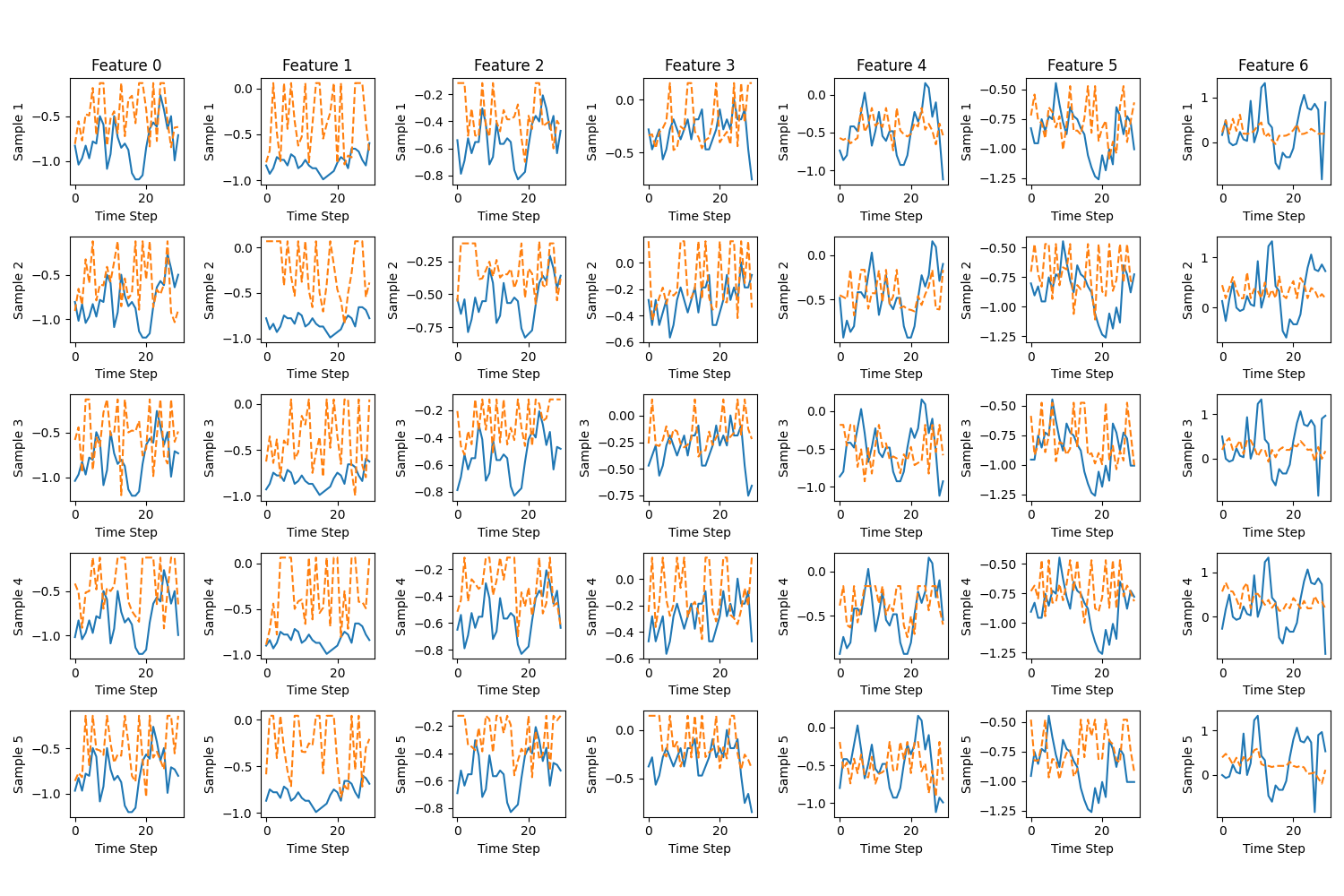}\\
    \caption{PatchTST prediction in the test dataset with the five test samples with the lowest MSE in the Air Quality Dataset}
    \label{fig9}
\end{figure}

\begin{figure}[H]
    \centering
    \includegraphics[width=1\textwidth]{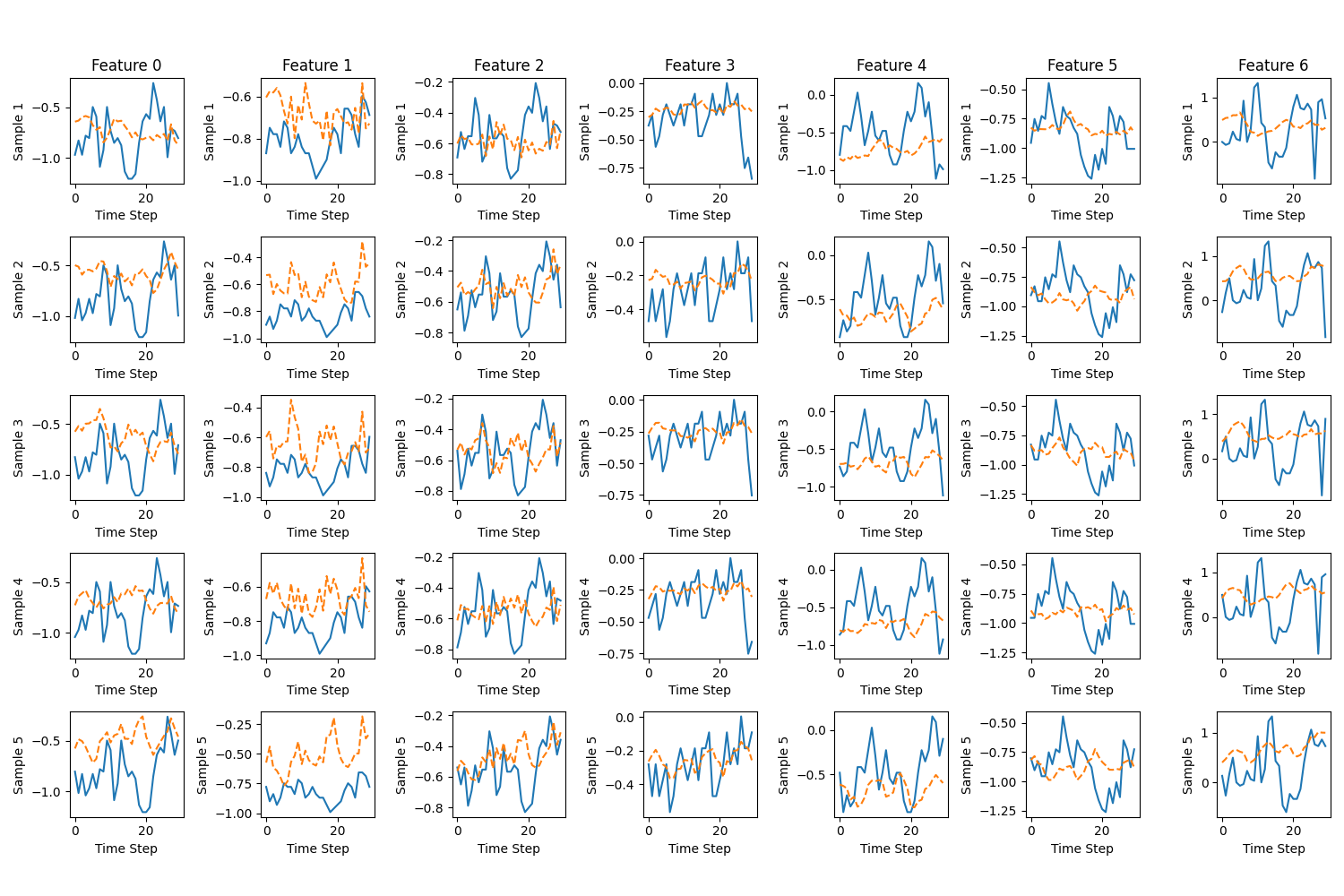}\\
    \caption{TimesNet prediction in the test dataset with the five test samples with the lowest MSE in the Air Quality Dataset}
    \label{fig10}
\end{figure}

\begin{figure}[H]
    \centering
    \includegraphics[width=1\textwidth]{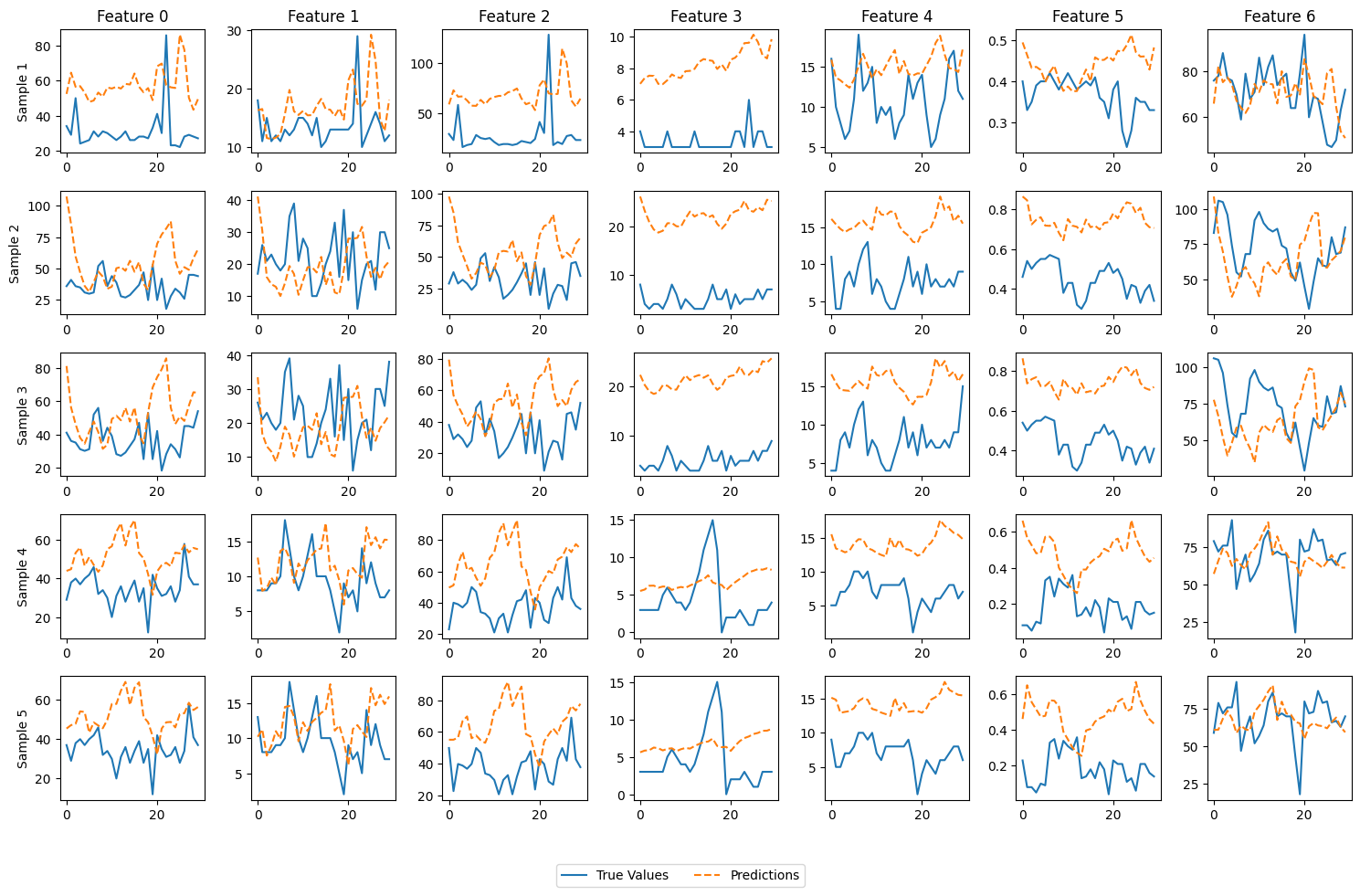}\\
    \caption{Transformer prediction in the test dataset with the five test samples with the lowest MSE in the Air Quality Dataset}
    \label{fig11}
\end{figure}

We visualized the predictions for seven features on five test samples with the lowest MSE to further examine the models' performance, as illustrated Figs.\ref{fig7}–\ref{fig11}. MMformer, as shown in Fig.\ref{fig7}, the prediction effect is significantly better than that of baseline models. It suggests MMformer's modeling structure and attention mechanisms more effectively capture global and local dependencies within the complex environmental MTS.

In contrast, Fig.\ref{fig8} demonstrates the challenges faced by the iTransformer model when dealing with rapid changes in the data, which illustrates how predictions become distorted under significant variability. For example, in several subplots, sudden peaks in the truth are smoothed or delayed in the predictions. It suggests that embedding timestamps alone may be insufficient to maintain sequence integrity across diverse regions and may introduce higher sensitivity to noise or regional variability.

Figs.\ref{fig9} and \ref{fig10} illustrate the prediction performance of PatchTST and TimesNet. These models effectively capture overall trends, with their predictions closely mirroring the actual data's trajectory. However, they tend to exhibit smoothing and lagging when faced with rapid changes, spikes, or sharp turns in the time series, suggesting limitations in capturing high-frequency dynamics and sudden temporal shifts. Fig.\ref{fig11} exhibits Transformer significant deviations from the truth, particularly in regions with abrupt fluctuations or sharp peaks. It suggests that the Transformer's self-attention may struggle to maintain accurate predictions during periods of high variability in MTS.

In conclusion, based on the air quality dataset, our combined quantitative evaluation and systematic graphical analysis illustrate the MMformer's superior performance. The observed shortcomings of the competing models in preserving critical patterns under adverse noise conditions further highlight the advantages of the proposed MMformer for modeling complex environmental MTS.

\paragraph{Comparative Analysis on the Climate Dataset}
In the climate dataset, MMformer will be compared with iTransformer, PatchTST, TimesNet, and Transformer using identical evaluation metrics. The analysis will assess actual prediction results and evaluation metrics to provide a comprehensive performance overview.

\begin{table}[ht]
    \centering
    \caption{Performance Comparison of Multivariate Time Series Forecasting Models on the Climate Dataset. The seq\_len and pred\_len are set to 40 and 10 for all models. The results are rounded to three decimal places. Figures in red and bold are the best for that metric.}
    \label{table8}
    \begin{threeparttable}
        \small
        \begin{tabularx}{\textwidth}{@{}>{\bfseries}lXXXXXXX@{}}
            \toprule
            \textbf{Model} & \textbf{MSE} & \textbf{MAE} & \textbf{MAPE}\\
            \midrule
            MMformer & \textcolor{red}{\textbf{0.283}} & \textcolor{red}{\textbf{0.237}} & \textcolor{red}{\textbf{13.623\%}}\\
            iTransformer & 0.578 & 0.342 & 18.878\%\\
            TimesNet & 0.593 & 0.362 & 20.171\%\\
            PatchTST & 0.634 & 0.388 & 21.304\%\\
            Transformer & 0.411 & 0.399 & 23.776\%\\
            \bottomrule
        \end{tabularx}
    \end{threeparttable}
\end{table}
Table \ref{table8} demonstrates that MMformer also significantly outperforms other strong baseline models on the climate dataset, achieving the lowest Mean Squared Error (MSE) of 0.283, Mean Absolute Error (MAE) of 0.237, and Mean Absolute Percentage Error (MAPE) of 13.623\%, which indicates excellent error minimization and data fitting capabilities. MMformer reduced MSE, MAE, and MAPE by 51.038\%, 30.702\%, and 27.837\%, respectively, compared to iTransformer; by 52.277\%, 34.530\%, and 32.463\% compared to TimesNet; and by 55.363\%, 38.918\%, and 36.054\% compared to PatchTST. Furthermore, MMformer achieved even greater performance gains against the Transformer model, with reductions of 31.144\% in MSE, 40.602\% in MAE, and 42.703\% in MAPE. 

The MMformer model's uncertainty quantification analysis on the climate dataset revealed a global average variance of 0.030902, corresponding to a global average standard deviation of 0.170853. It indicates the overall dispersion of the model's predictions. The variance across all samples, timesteps, and features ranged from a minimum of 0.007938 to a maximum of 0.147739. Using a 95\% confidence level (with a Z-score of approximately 1.96), the average width of the prediction confidence interval can be calculated as: $2 \times 1.96 \times 0.170853 \approx 0.6690$. It allows for the evaluation of predictive confidence and the assessment of potential error margins in decision-making processes.

\begin{figure}[H]
    \centering
    \includegraphics[width=1\textwidth]{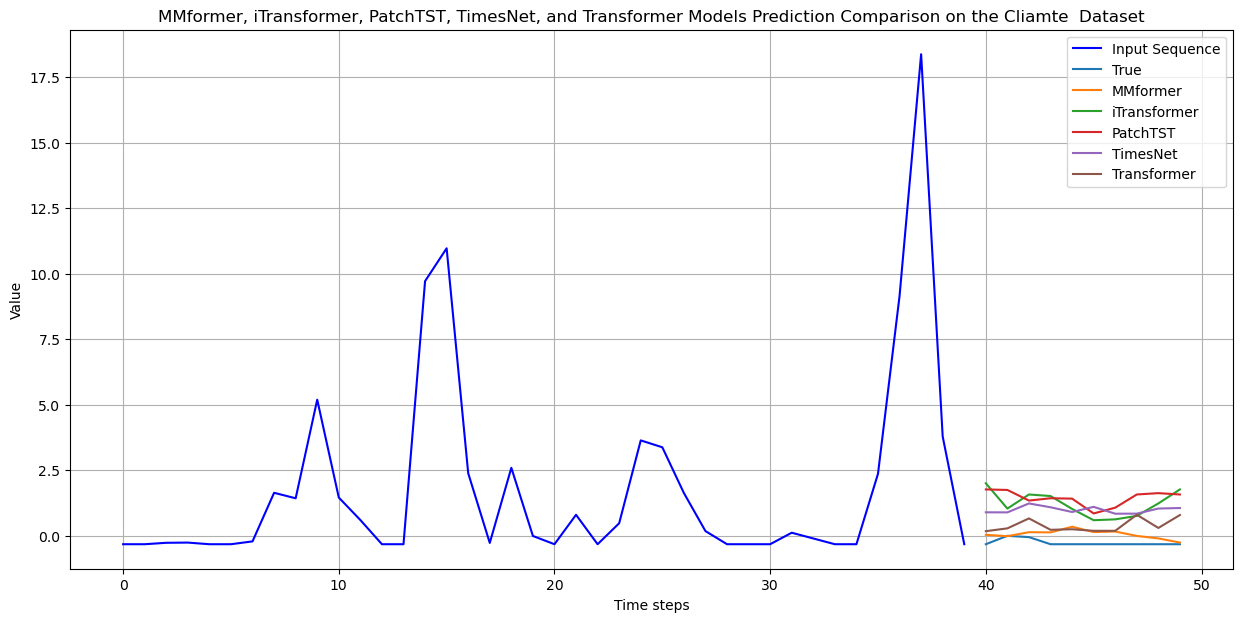}\\
    \caption{MMformer, iTransformer, PatchTST, TimesNet, and Transformer Models Prediction Comparison on the Cliamte Dataset}
    \label{fig12}
\end{figure}

The dataset is partitioned into 1461 days for training, 181 days for validation, and the remaining days for testing, with sliding window techniques employed to maximize training data utilization. Fig.\ref{fig12} presents a prediction comparison of baseline models using a sample from the climate dataset. MMformer successfully captures these subsequent low actual values, thereby validating its adaptive capacity to varying temporal dynamics. The visual comparison further reinforces the quantitative metrics reported earlier, collectively demonstrating the comprehensive superiority of MMformer over the other evaluated models. The climate dataset, comprising five years (2018-2022, 1826 days) of continuous rainfall and temperature data, is selected for its lack of strong seasonal bias, which clearly demonstrates temporal variation robustness.

\begin{figure}[H]
    \centering
    \includegraphics[width=1\textwidth]{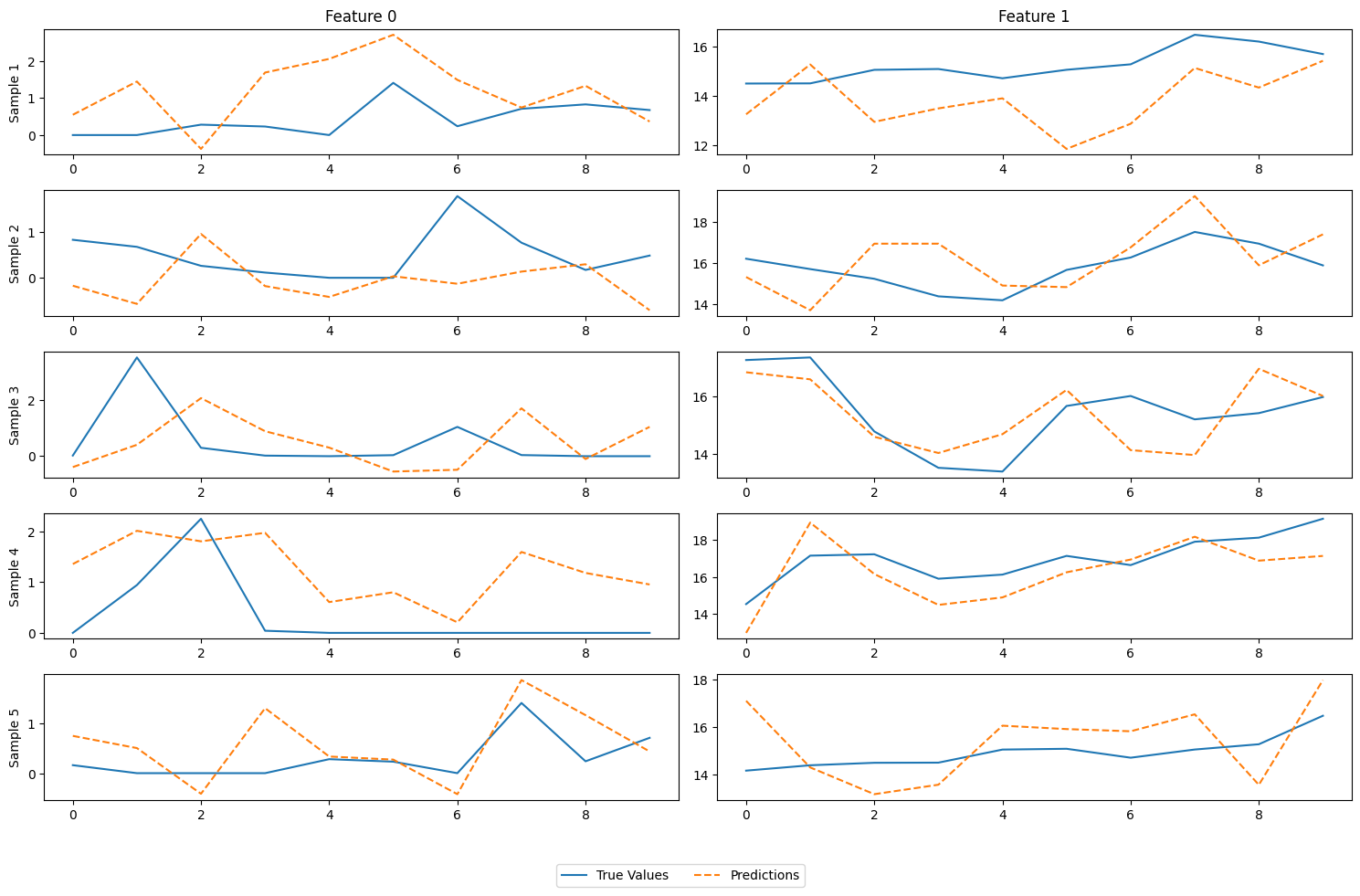}\\
    \caption{MMformer prediction in the test dataset with the five test samples with the lowest MSE in the Climate Dataset}
    \label{fig13}
\end{figure}

\begin{figure}[H]
    \centering
    \includegraphics[width=1\textwidth]{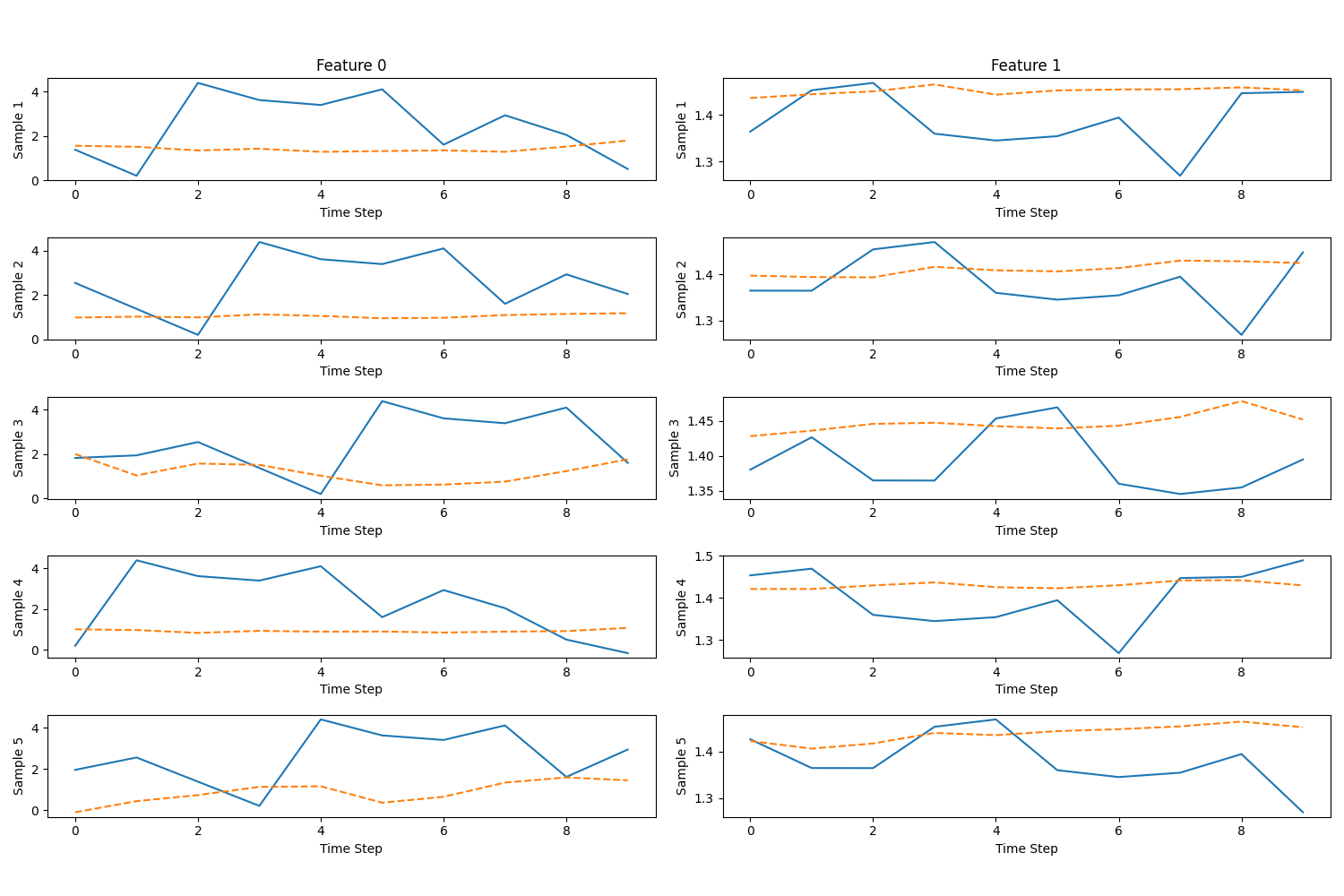}\\
    \caption{iTransformer prediction in the test dataset with the five test samples with the lowest MSE in the Climate Dataset}
    \label{fig14}
\end{figure}

\begin{figure}[H]
    \centering
    \includegraphics[width=1\textwidth]{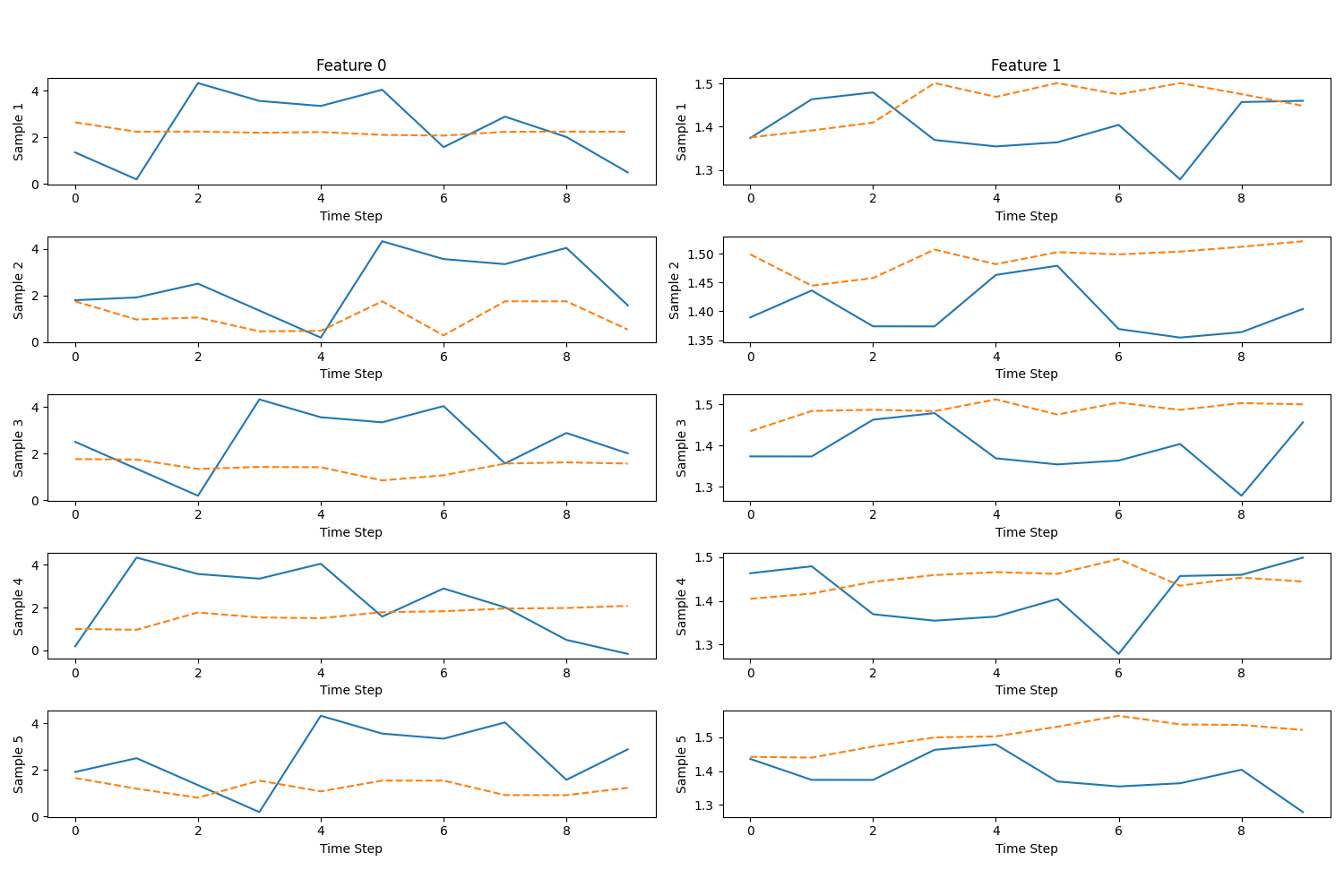}\\
    \caption{PatchTST prediction in the test dataset with the five test samples with the lowest MSE in the Climate Dataset}
    \label{fig15}
\end{figure}

\begin{figure}[H]
    \centering
    \includegraphics[width=1\textwidth]{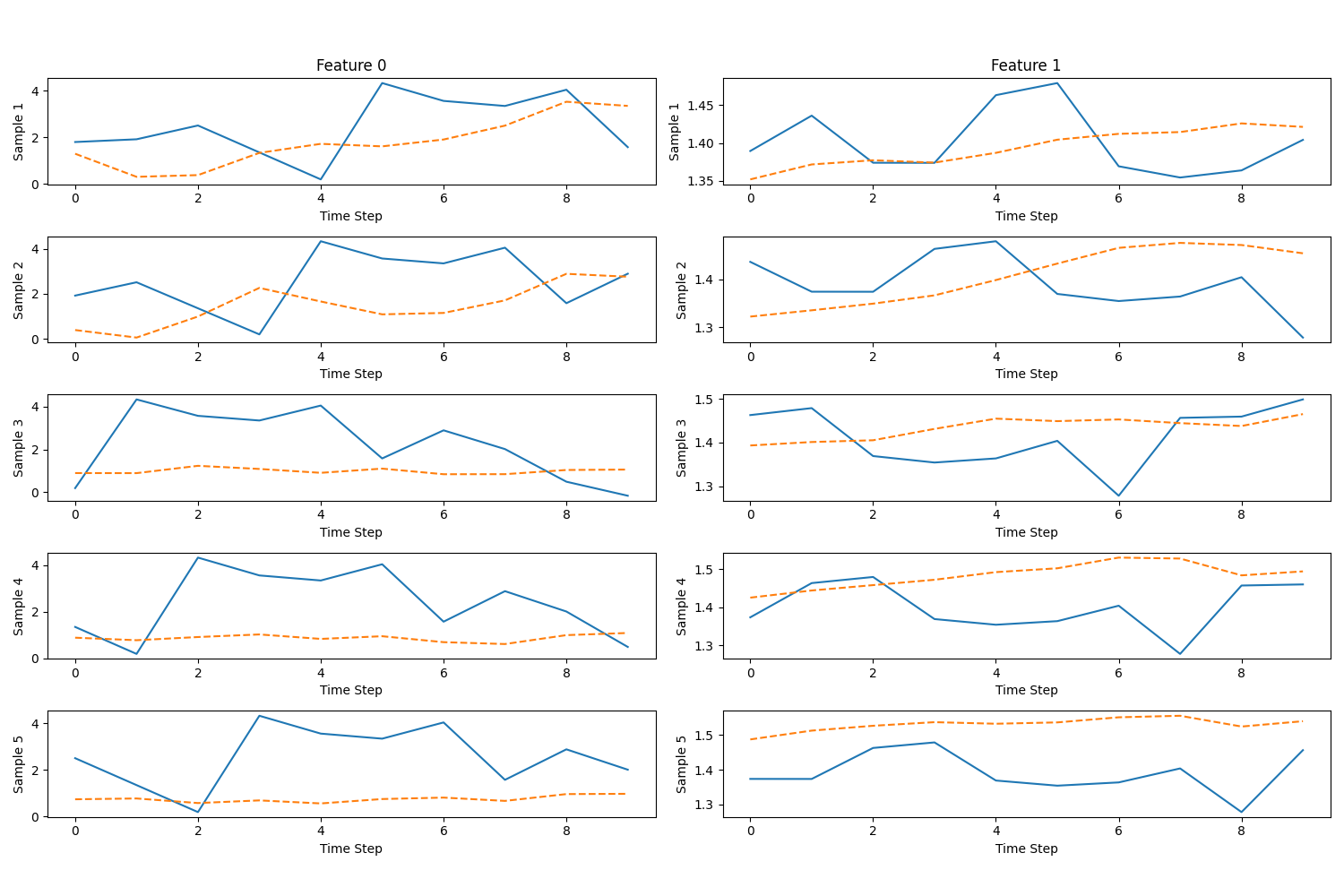}\\
    \caption{TimesNet prediction in the test dataset with the five test samples with the lowest MSE in the Climate Dataset}
    \label{fig16}
\end{figure}

\begin{figure}[H]
    \centering
    \includegraphics[width=1\textwidth]{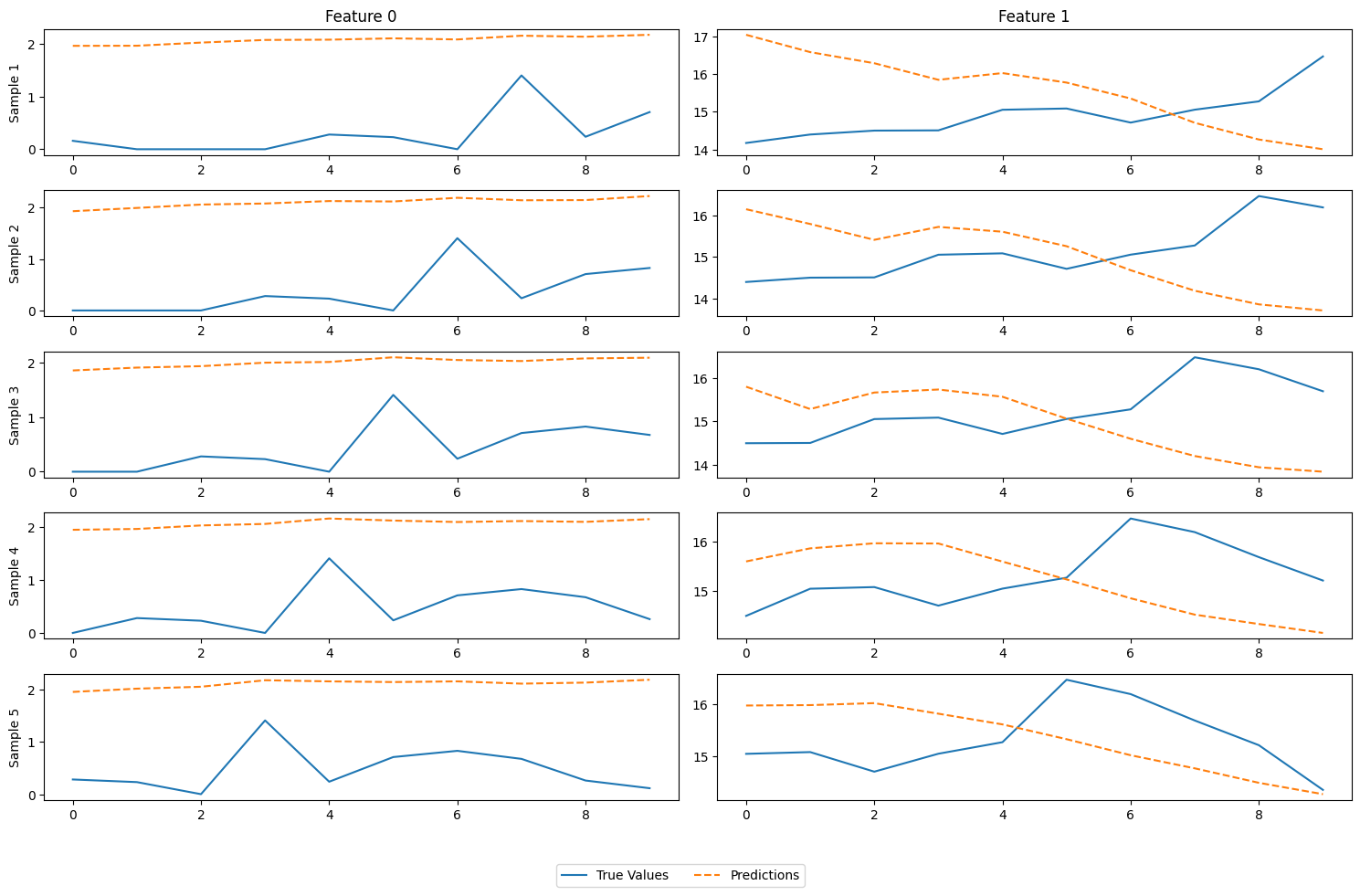}\\
    \caption{Transformer prediction in the test dataset with the five test samples with the lowest MSE in the Climate Dataset}
    \label{fig17}
\end{figure}

To further assess model performance, we visualized the predictions for two features from the five test samples with the lowest MSE (Figs.\ref{fig13}-\ref{fig17}). As shown in Fig.\ref{fig13}, MMformer accurately captures the overall trends and relationships between the selected features. The close alignment between predicted and actual values indicates that MMformer effectively maintains global dependencies between features, even when the dimensionality of the dataset is reduced.


In contrast, Figs.\ref{fig14}-\ref{fig17} reveal limitations in the ability of iTransformer, PatchTST, TimesNet, and Transformer to model the local interactions between the two features jointly. While these models perform well on the dataset according to the metrics in Table \ref{table8}, the visualizations show that they struggle to capture the nuanced, localized patterns in the authentic feature relationships. The predicted values often deviate significantly from the ground truth, suggesting these models fail to capture complex environmental MTS local features.

By comparing the performance of these baseline models on metrics and visualization plots, it is clear that the MMformer is more effective at learning global trends and local interactions between features on real-world data. Whether it is an air quality dataset or a climate dataset, our combined quantitative evaluation and systematic graphical analysis provide compelling evidence of the MMformer's superior performance compared to the iTransformer, PatchTST, TimesNet, and Transformer strong baseline models on real-world environmental datasets. To further evaluate MMformer's generalization capabilities, we applied it to the widely adopted MTS datasets, PEMS.

\paragraph{Comparative Analysis of Prediction Methods in Widely Used MTS Datasets}
To further validate MMformer's generalization capabilities across diverse MTS benchmarks, we conducted a comparative analysis on the widely used PEMS datasets. The evaluation utilized four PEMS datasets (PEMS03, PEMS04, PEMS07, and PEMS08), with a sequence length of 40 and a prediction length of 10. MMformer is benchmarked against iTransformer, TimesNet, PatchTST, and Transformer, employing MSE, MAE, and MAPE as key evaluation metrics.

\begin{table}[H]
    \centering
    \caption{Performance Comparison of Multivariate Time Series Forecasting Models on PEMS Datasets. The seq\_len and pred\_len are set to 40 and 10 for all models. The results are rounded to three decimal places. Figures in red and bold are the best for that metric, blue and bold indicate that metric is second.}
    \label{table9}
    \adjustbox{width=\textwidth,center}
    {
        \small
        \begin{tabular}{l|*{5}{cc}}
            \toprule
            \multirow{2}{*}{\textbf{Models}} & \multicolumn{2}{c|}{\textbf{MMformer}} & \multicolumn{2}{c|}{\textbf{iTransformer}} & \multicolumn{2}{c|}{\textbf{TimesNet}} & \multicolumn{2}{c|}{\textbf{PatchTST}} & \multicolumn{2}{c}{\textbf{Transformer}}\\
            & \multicolumn{2}{c|}{\textbf{(Ours)}} & \multicolumn{2}{c|}{\textbf{(2024)}} & \multicolumn{2}{c|}{\textbf{(2023)}} & \multicolumn{2}{c|}{\textbf{(2023)}} & \multicolumn{2}{c}{\textbf{(2017)}} \\
            \cline{2-11}
            \textbf{Metric} & \textbf{MSE} & \textbf{MAE} & \textbf{MSE} & \textbf{MAE} & \textbf{MSE} & \textbf{MAE} & \textbf{MSE} & \textbf{MAE} & \textbf{MSE} & \textbf{MAE} \\
            \hline
            PEMS03 & \textcolor{red}{\textbf{0.066}} & \textcolor{red}{\textbf{0.173}} & 
            0.080 & 0.189 & 
            \textcolor{blue}{\textbf{0.075}} & \textcolor{blue}{\textbf{0.179}} & 
            0.100 & 0.219 & 
            0.116 & 0.226 \\
            PEMS04 & \textcolor{red}{\textbf{0.201}} & \textcolor{red}{\textbf{0.225}} & 
            \textcolor{blue}{\textbf{0.255}} & \textcolor{blue}{\textbf{0.252}} & 
            0.298 & 0.273 & 
            0.265 & 0.271 & 
            0.332 & 0.320 \\
            PEMS07 & 0.075 & 0.195 & 
            \textcolor{red}{\textbf{0.067}} & \textcolor{red}{\textbf{0.168}} & 
            \textcolor{blue}{\textbf{0.075}} & \textcolor{blue}{\textbf{0.175}} & 
            0.091 & 0.210 & 
            0.166 & 0.250 \\
            PEMS08 & \textcolor{red}{\textbf{0.196}} & \textcolor{red}{\textbf{0.226}} & 
            \textcolor{blue}{\textbf{0.270}} & \textcolor{blue}{\textbf{0.253}} & 
            0.420 & 0.288 & 
            0.298 & 0.283 & 
            0.642 & 0.386 \\
            \bottomrule
        \end{tabular}
    }
\end{table}
\begin{table}[H]
    \centering
    \caption{Comparison of MAPE for Multivariate Time Series Forecasting Models on PEMS Datasets. The seq\_len and pred\_len are set to 40 and 10 for all models. The results are rounded to three decimal places. Figures in red and bold are the best for that metric, blue and bold indicate that metric is second.}
    \label{table10}
    \adjustbox{width=\textwidth,center}
    {
        \small
        \begin{tabular}{l|cccccc}
            \hline
            \textbf{Models} & \textbf{MMformer} & \textbf{iTransformer} & \textbf{TimesNet} & \textbf{PatchTST} & \textbf{Transformer}\\
            \textbf{Metric} & \textbf{MAPE} & \textbf{MAPE} & \textbf{MAPE} & \textbf{MAPE} & \textbf{MAPE}\\
            \hline
            PEMS03 & \textcolor{red}{\textbf{9.801\%}} & 11.318\% & \textcolor{blue}{\textbf{10.706\%}} & 13.057\% & 12.997\%\\
            PEMS04 & \textcolor{red}{\textbf{12.803\%}} & \textcolor{blue}{\textbf{14.496\%}} 
            & 15.584\% & 15.494\% & 17.706\%\\
            PEMS07 & 11.428\% & \textcolor{red}{\textbf{10.131\%}} & \textcolor{blue}{\textbf{10.524\%}} & 12.605\% & 14.055\%\\
            PEMS08 & \textcolor{red}{\textbf{12.684\%}} & \textcolor{blue}{\textbf{14.341\%}} & 16.138\% & 16.100\% & 19.908\%\\
            \hline
        \end{tabular}
    }
\end{table}
\begin{table}[H]
    \centering
    \caption{Performance Comparison of Forecasting Baselines against the MMformer on PEMS Datasets. The seq\_len and pred\_len are set to 40 and 10 for all models. Results are rounded to three decimal places. Figures in blue and bold indicate MMformer is inferior to the model in this indicator.}
    \label{table11}
    \adjustbox{width=\textwidth,center}
    {
        \small
        \begin{tabular}{l|*{4}{cc}}
            \toprule
            \multirow{2}{*}{\makecell{\textbf{Models} \\ \textbf{Metric}}} & \multicolumn{2}{c|}{\textbf{iTransformer (2024)}} & \multicolumn{2}{c|}{\textbf{TimesNet (2023)}} & \multicolumn{2}{c|}{\textbf{PatchTST (2023)}} & \multicolumn{2}{c}{\textbf{Transformer (2017)}}\\
            \cline{2-9}
            & \textbf{MSE} & \textbf{MAE} & \textbf{MSE} & \textbf{MAE} & \textbf{MSE} & \textbf{MAE} & \textbf{MSE} & \textbf{MAE} \\
            \hline
            PEMS03 &  +21.212\% & +9.249\%  & +13.636\% & +3.468\% & +51.515\% & +26.590\% & +75.758\% & +30.636\%  \\
            PEMS04 & +26.866\% & +12.000\%  & +48.259\% & +21.333\% & +31.841\% & +20.444\% & +65.174\% & +42.222\%  \\
            PEMS07 & \textcolor{blue}{\textbf{-10.667\%}}  & \textcolor{blue}{\textbf{-13.846\%}}  & 0.000\% & \textcolor{blue}{\textbf{-10.256\%}}  & +21.333\% & +7.692\%  & +121.333\%  & +28.205\%   \\
            PEMS08 & +37.755\%  & +11.947\%  & +114.286\% & +27.434\%  & +52.041\% & +25.221\% & +229.592\%  & +70.796\%   \\
            \bottomrule
        \end{tabular}
    }
\end{table}
The experimental results, as detailed in Tables \ref{table9} through \ref{table11}, indicate that MMformer consistently delivers superior performance across the evaluated PEMS datasets. The evaluation focuses on Mean Squared Error (MSE), Mean Absolute Error (MAE), and Mean Absolute Percentage Error (MAPE).

On the PEMS03 dataset, MMformer exhibited the best performance, achieving the lowest MSE (0.066), MAE (0.173), and MAPE (9.801\%). TimesNet followed, with an MSE of 0.075, MAE of 0.179, and MAPE of 10.706\%. Compared to MMformer, iTransformer's MSE and MAE increased by 21.212\% and 9.249\%, respectively. PatchTST shows increases of 51.515\% in MSE and 26.590\% in MAE, while Transformer's MSE and MAE increased by 75.758\% and 30.636\%, respectively.

For the PEMS04 dataset, MMformer again demonstrated superior performance, obtaining the best MSE (0.201), MAE (0.225), and MAPE (12.803\%). iTransformer ranked second with an MSE of 0.255, MAE of 0.252, and MAPE of 14.496\%. Compared to MMformer, iTransformer’s MSE and MAE increased by 26.866\% and 12.000\%, respectively. PatchTST's MSE and MAE increased by 31.841\% and 20.444\%, while Transformer’s MSE and MAE increased by 65.174\% and 42.222\%, respectively.

In the PEMS07 evaluation, iTransformer led in MSE (0.067), MAE (0.168), and MAPE (10.131\%). MMformer also performed well, ranking third with an MSE of 0.075, MAE of 0.195, and MAPE of 11.428\%. TimesNet secured the second position in MSE (0.075), MAE (0.175), and MAPE (10.524\%). Compared to iTransformer, MMformer’s MSE and MAE increased by 10.667\% and 13.846\%, respectively. TimesNet's MSE is comparable to MMformer's, with an MAE increase of 10.256

MMformer’s performance is also strong on the PEMS08 dataset, achieving the best MSE (0.196), MAE (0.226), and MAPE (12.684\%). iTransformer ranked second with an MSE of 0.270, MAE of 0.253, and MAPE of 14.341\%. Compared to MMformer, iTransformer's MSE and MAE increased by 37.755\% and 11.947\%, respectively. PatchTST’s MSE and MAE increased by 52.041\% and 25.221\%, while Transformer’s MSE and MAE increased by 229.592\% and 70.796\%, respectively.

Overall, MMformer demonstrated a consistently strong performance across the PEMS datasets. The experiments validate that MMformer is a viable and robust model for multivariate time series forecasting tasks, achieving SOTA or near-SOTA performance across multiple datasets and metrics. While iTransformer performs exceptionally well on the PEMS07 dataset, MMformer performs better on the other datasets. The highlights MMformer’s effectiveness and robustness in handling high-dimensional and complex MTS forecasting tasks, thus showcasing its substantial potential for environmental applications and other domains with similar data characteristics.

\subsubsection{Ablation Study}
To assess the contribution of each component in MMformer, we conducted ablation experiments by individually removing the MAML and MC Dropout modules. The resulting models are evaluated using MSE, MAE, and MAPE metrics on the air quality dataset.

\begin{table}[H]
    \centering
    \caption{MMformer Ablation Experiment Results Using a Real-World Air Quality Dataset. The seq\_len and pred\_len are set to 150 and 30 for all models. The results are rounded to three decimal places. Figures in red and bold are the best for that metric.}
    \label{table5}
    \begin{threeparttable}
        \small
        \begin{tabularx}{\textwidth}{@{}>{\bfseries}lXXXXXXX@{}}
            \toprule
            \textbf{Model} & \textbf{MSE} & \textbf{MAE} & \textbf{MAPE}\\
            \midrule
            MMformer & 0.245 & \textcolor{red}{\textbf{0.332}} & \textcolor{red}{\textbf{18.873\%}}\\
            Remove MAML & 0.252 & 0.338 & 19.166\%\\
            Remove MC Dropout & \textcolor{red}{\textbf{0.235}} & 0.338 & 19.555\%\\
            Remove MAMAL and MC Dropout & 0.239 & 0.334 & 19.323\%\\
            \bottomrule
        \end{tabularx}
    \end{threeparttable}
\end{table}
Table \ref{table5} summarizes the results of the ablation study, comparing the performance of the original MMformer model with three modified versions: MMformer without meta-learning (MAML), MMformer without MC Dropout, and MMformer without both meta-learning and MC Dropout. As shown in Table \ref{table5}, the original MMformer achieves the lowest MSE (0.245), MAE (0.332), and MAPE (18.873\%), indicating the highest predictive accuracy among all configurations. Removing meta-learning (MAML) results in a significant decline in performance, with MSE increasing to 0.252, MAE to 0.338, and MAPE to 19.166\%. Excluding MC Dropout also leads to reduced performance, though the impact is less pronounced (MSE reducing to 0.235, but MAE: 0.338, MAPE: 19.555\%). The removal of both meta-learning and MC Dropout causes the MSE to decrease to 0.239, but the MAE rises to 0.334, and the MAPE to 19.323\%. These findings demonstrate that meta-learning and MC Dropout are critical for improving the accuracy and robustness of the MMformer model.

Ablation studies reveal that MC Dropout negatively impacts MSE. However, its crucial role in providing uncertainty analysis for climate and environmental applications, alongside its positive effects on MAE and MAPE, collectively enhances the model's overall performance.

\begin{figure}[H]
    \centering
    \includegraphics[width=1\textwidth]{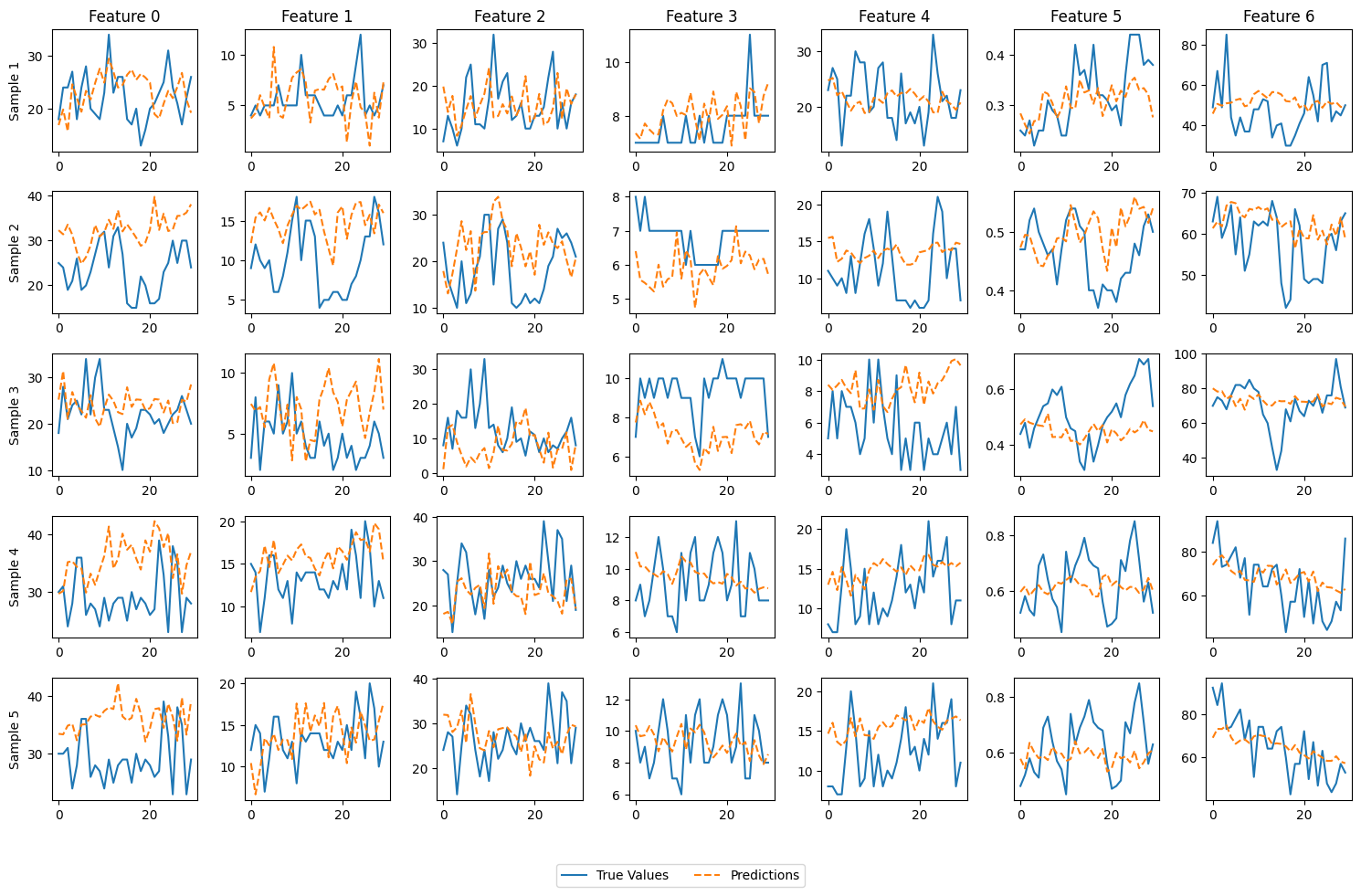}\\
    \caption{MMformer -remove MAML Prediction in Test Dataset with the 5 test samples with the lowest MSE}
    \label{fig18}
\end{figure}

\begin{figure}[ht]
    \centering
    \includegraphics[width=1\textwidth]{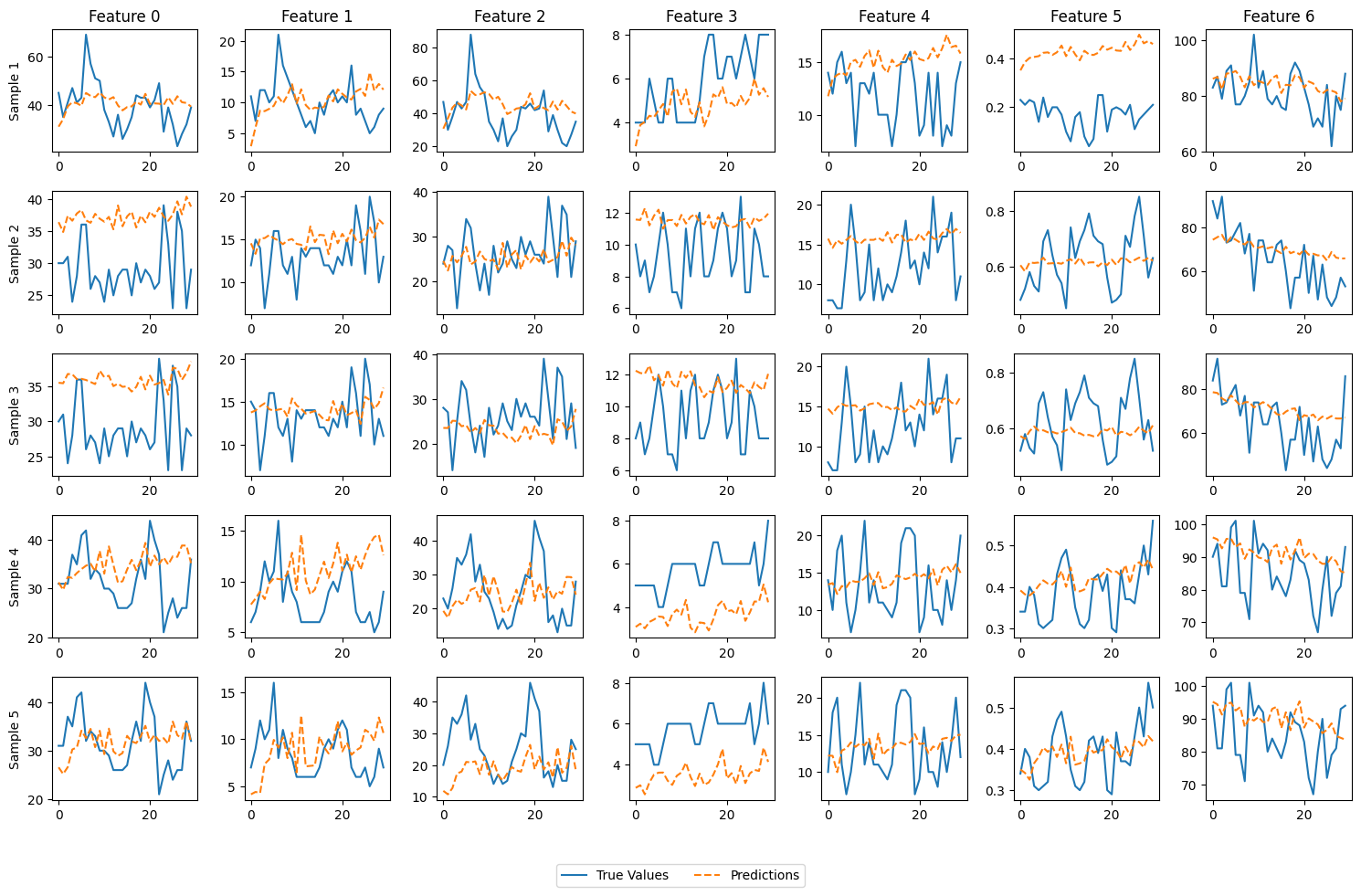}\\
    \caption{MMformer -remove MC Dropout Prediction in Test Dataset with the 5 test samples with the lowest MSE}
    \label{fig19}
\end{figure}

\begin{figure}[H]
    \centering
    \includegraphics[width=1\textwidth]{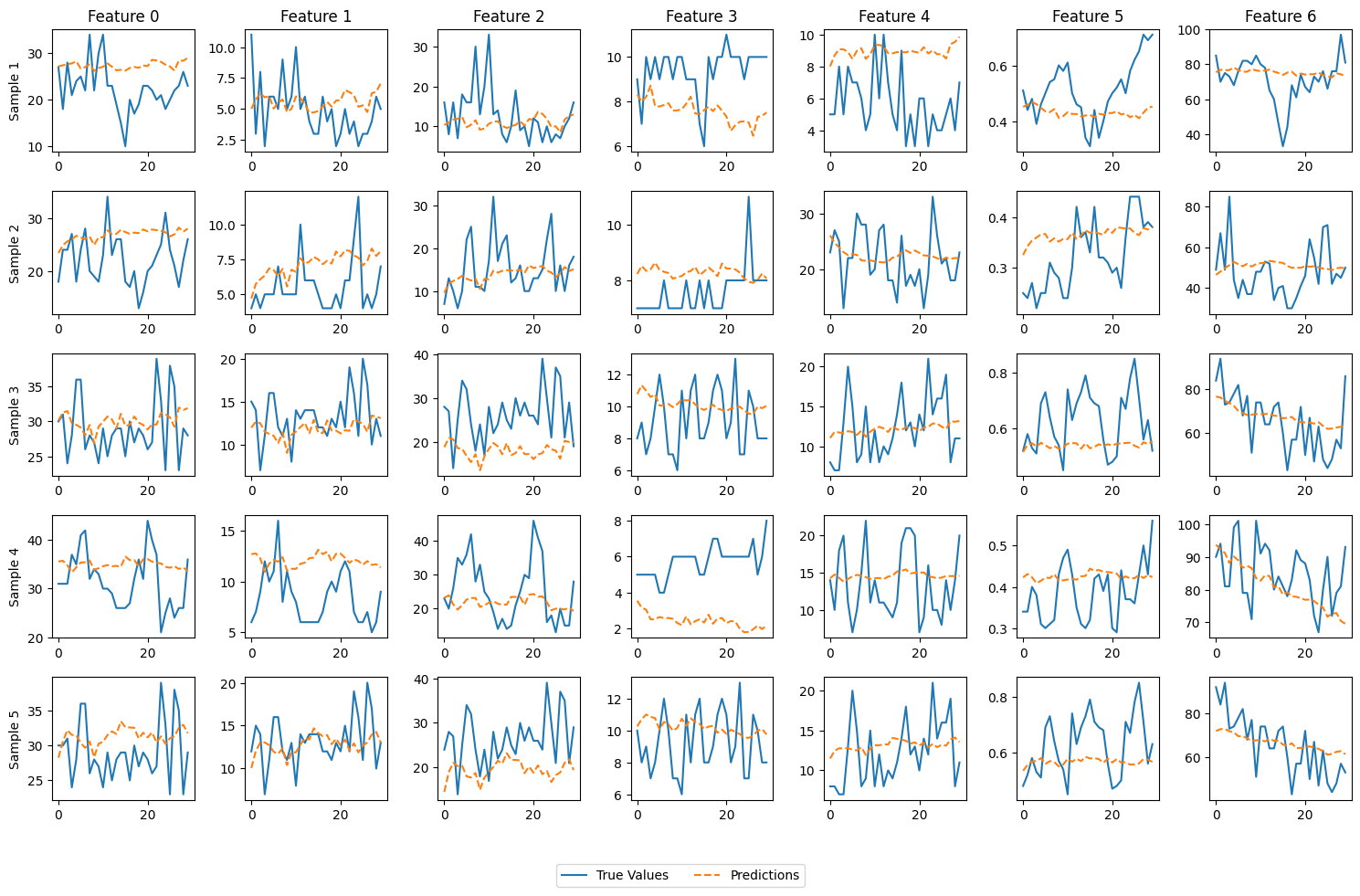}\\
    \caption{MMformer -remove MAML and MC Dropout Prediction in Test Dataset with the 5 test samples with the lowest MSE}
    \label{fig20}
\end{figure}

Figs.\ref{fig18}-\ref{fig20} offer a more detailed visualization of the predicted trajectories under the various ablation settings. A direct comparison of those figures demonstrates that excluding MAML or MC Dropout leads to more pronounced deviations from the ground truth, particularly when abrupt changes occur in the time series. Specifically, excerpts where the data fluctuates rapidly highlight lagged or oversmoothed predictions for the ablated variants, whereas the complete MMformer maintains closer alignment with observed values.

Ablation experiment evaluation quantitative outcomes (Table \ref{table5}) and visual analysis (Figs.\ref{fig18}-\ref{fig20}) demonstrate that both MAML and MC Dropout play a critical role in enhancing the MMformer’s capability to learn complex temporal dynamics, thereby ensuring consistent and robust performance in multivariate time-series forecasting. The model's multi-layer linear structure and MAML further help avoid overfitting. Ablation experiments, which involved removing these modules, revealed notable decreases in prediction accuracy and robustness.

\subsubsection{Robustness and Efficiency Analysis}
To comprehensively evaluate MMformer's practical applicability, we conducted a thorough analysis of its robustness and computational efficiency. Robustness is assessed by examining performance stability under varying pred\_len, while efficiency is measured through computational resource consumption and time during training and inference.

A robust forecasting model should maintain high performance not only across different prediction tasks but also show low sensitivity to its internal hyperparameter settings. As shown in Table \ref{table12}, we test the robustness of the model to different forecasting horizons on the Air Quality dataset with prediction lengths (pred\_len) set to 30, 60, and 90. Across the same seq\_len (150), d\_model (768), n\_heads(8), encoder\_layers (6), d\_ff (2048), dropout (0,2), MMformer consistently achieved the lowest MSE and MAE, demonstrating its superior stability and accuracy in handling both short-term and long-term forecasting challenges. While other models experienced more significant performance degradation as the prediction length increased, MMformer maintained a clear advantage, underscoring the robustness of its architecture in capturing temporal dependencies over extended horizons. MMformer's low sensitivity suggests that MMformer does not require exhaustive hyperparameter tuning to achieve its strong results, confirming its robustness and ease of deployment.
\begin{table}[H]
    \centering
    \caption{Robustness Comparison of Multivariate Time Series Forecasting Models on Air Quality Datasets. Values are presented to three decimal places, with optimal results for each metric in bold red and the second in bold blue. (First label column split into Dataset and Prediction Length)}
    \label{table12}
    \adjustbox{width=\textwidth,center}
    {
        \small
        \begin{tabular}{c |c | cc | cc | cc | cc | cc}
            \toprule

            \multirow{2}{*}{\textbf{Models}} & \multicolumn{1}{c}{\textbf{}} &
            \multicolumn{2}{c}{\textbf{MMformer}} & 
            \multicolumn{2}{c}{\textbf{iTransformer}} & 
            \multicolumn{2}{c}{\textbf{TimesNet}} & 
            \multicolumn{2}{c}{\textbf{PatchTST}} & 
            \multicolumn{2}{c}{\textbf{Transformer}}\\
            
            & \multicolumn{1}{c}{\textbf{}} &\multicolumn{2}{c}{\textbf{(Ours)}} & 
            \multicolumn{2}{c}{\textbf{(2024)}} & 
            \multicolumn{2}{c}{\textbf{(2023)}} & 
            \multicolumn{2}{c}{\textbf{(2023)}} & 
            \multicolumn{2}{c}{\textbf{(2017)}} \\
            \cline{2-12}
            \multirow{1}{*}{\textbf{Metric}} & \multicolumn{1}{c}{\textbf{}} & \textbf{MSE} & \textbf{MAE} & 
            \textbf{MSE} & \textbf{MAE} & 
            \textbf{MSE} & \textbf{MAE} & 
            \textbf{MSE} & \textbf{MAE} & 
            \textbf{MSE} & \textbf{MAE} \\ 
            \midrule 
            
            \multirow{5}{*}{\rotatebox{90}{\textbf{Air Quality}}} &
            \textbf{30} & \textcolor{red}{\textbf{0.245}} & \textcolor{red}{\textbf{0.332}} & 
            0.822 & 0.529 & 
            0.807 & 0.528 & 
            0.862 &  0.556 & 
            \textcolor{blue}{\textbf{0.770}} & \textcolor{blue}{\textbf{0.524}} \\

            \textbf{} & \textbf{60} & \textcolor{red}{\textbf{0.779}} & \textcolor{red}{\textbf{0.472}} & 
            0.901 & 0.565 & 
            0.875 & \textcolor{blue}{\textbf{0.552}} & 
            0.957 & 0.602 & 
            \textcolor{blue}{\textbf{0.835}} & 0.556 \\
            
            \textbf{} & \textbf{90} & \textcolor{red}{\textbf{0.844}} & \textcolor{red}{\textbf{0.491}} & 
            0.977 & 0.594 & 
            0.944 & 0.587 & 
            1.016 & 0.626 & 
            \textcolor{blue}{\textbf{0.898}} & \textcolor{blue}{\textbf{0.577}} \\
            
            \cmidrule(lr){3-12}
            
            \textbf{} & \textbf{Avg} & \textcolor{red}{\textbf{0.623}} & \textcolor{red}{\textbf{0.432}} & 
            0.900 & 0.563 & 
            0.875 & 0.556 & 
            0.945 & 0.595 & 
            \textcolor{blue}{\textbf{0.834}} & \textcolor{blue}{\textbf{0.552}} \\
            \bottomrule
        \end{tabular}
    }
\end{table}

\begin{table}[H]
    \centering
    \caption{Efficiency Comparison of Multivariate Time Series Forecasting Models on Air Quality Datasets. Values are presented to three decimal places, with optimal results for each metric in bold red and the second in bold blue. (First label column split into Dataset and Prediction Length)}
    \label{table13}
    \adjustbox{width=\textwidth,center}
    {
        \small
        \begin{tabular}{c |c | cc | cc | cc | cc | cc}
            \toprule

            \multirow{2}{*}{\textbf{Models}} & \multicolumn{1}{c}{\textbf{}} &
            \multicolumn{2}{c}{\textbf{MMformer}} & 
            \multicolumn{2}{c}{\textbf{iTransformer}} & 
            \multicolumn{2}{c}{\textbf{TimesNet}} & 
            \multicolumn{2}{c}{\textbf{PatchTST}} & 
            \multicolumn{2}{c}{\textbf{Transformer}}\\
            
            & \multicolumn{1}{c}{\textbf{}} &\multicolumn{2}{c}{\textbf{(Ours)}} & 
            \multicolumn{2}{c}{\textbf{(2024)}} & 
            \multicolumn{2}{c}{\textbf{(2023)}} & 
            \multicolumn{2}{c}{\textbf{(2023)}} & 
            \multicolumn{2}{c}{\textbf{(2017)}} \\
            \cline{2-12}
            \multirow{1}{*}{\textbf{Metric}} & \multicolumn{1}{c}{\textbf{}} & \textbf{Train} & \textbf{Inference} & 
            \textbf{Train} & \textbf{Inference} &
            \textbf{Train} & \textbf{Inference} &
            \textbf{Train} & \textbf{Inference} &
            \textbf{Train} & \textbf{Inference} \\
            \midrule 
            
            \multirow{5}{*}{\rotatebox{90}{\textbf{Air Quality}}} &
            \textbf{30} & \textcolor{blue}{\textbf{3796}} & 621 & 
            \textcolor{red}{\textbf{3597}} & 924 & 
            16581 & \textcolor{blue}{\textbf{421}} & 
            11837 &  \textcolor{red}{\textbf{191}} & 
            18567 & 2623 \\

            \textbf{} & \textbf{60} & \textcolor{red}{\textbf{2859}} & \textcolor{blue}{\textbf{414}} & 
            \textcolor{blue}{\textbf{5053}} & 1610 & 
            19819 & 492 & 
            6769 & \textcolor{red}{\textbf{192}} & 
            20176 & 2878 \\
            
            \textbf{} & \textbf{90} & \textcolor{red}{\textbf{2783}} & \textcolor{blue}{\textbf{218}} & 
            \textcolor{blue}{\textbf{4108}} & 1668 & 
            33677 & 543 & 
            6755 & \textcolor{red}{\textbf{198}} & 
            20973 & \textcolor{blue}{\textbf{2844}} \\
            
            \cmidrule(lr){3-12}
            
            \textbf{} & \textbf{Avg} & \textcolor{red}{\textbf{3146}} & \textcolor{blue}{\textbf{418}} & 
            \textcolor{blue}{\textbf{4253}} & 1401 & 
            23359 & 485 & 
            8454 & \textcolor{red}{\textbf{194}} & 
            19905 & 2782 \\
            \bottomrule
        \end{tabular}
    }
\end{table}

For a model to be practical, especially for large-scale environmental monitoring, it must be computationally efficient. We compared MMformer's efficiency against baselines by measuring training time, inference time, total parameter count, and peak memory usage. All experiments are conducted on the same hardware (Intel Core i9-13900kf CPU, 32GB RAM, NVIDIA RTX 3090 GPU) to ensure a fair comparison.

Table \ref{table13} presents a comprehensive comparison of model efficiency. The analysis highlights MMformer's competitive training and inference speeds. For instance, in the most common forecasting setting (pred\_len=30), MMformer's training time is significantly lower than that of TimesNet and Transformer, and comparable to the highly efficient iTransformer. While PatchTST shows the fastest inference time, MMformer's inference speed is practical for most real-world applications and substantially faster than other baselines.

In terms of model complexity, MMformer strikes an effective balance. The parameter count of MMformer is 26, which is comparable to the baseline models' parameter count of around 20. During training, the peak memory usage of MMformer and TimesNet is measured at 21 GB. In contrast, models like iTransformer and Transformer of memory usage are 7 GB, and the memory usage of PatchTST is 5 GB. While MMformer is memory-intensive during the training phase, it is highly efficient in deployment, characterized by rapid inference speeds and a minimal memory footprint. The Tables \ref{table12} and \ref{table13} solidify MMformer's case as a high-performance yet practical solution for real-world multivariate time series forecasting.

\section{Discussion}
This study introduces MMformer to address critical challenges in multi-regional MTS prediction, specifically model adaptability, cross-task generalization, and uncertainty quantification. Traditional methods and earlier deep learning models like RNNs remain inadequately addressed in these aspects. MMformer enhances predictive accuracy and complex spatiotemporal time series modeling by optimizing an Encoder-only architecture through integrated timestamp handling, dimension inversion, the ATMA mechanism, and MC Dropout. The core innovation lies in the synergistic combination of these components, which effectively tackles MTS prediction issues, including data heterogeneity, prediction uncertainty, and the specific demands of multi-regional multiple features applications, delivering a more robust and adaptive solution.

\subsection{Findings Relative to Research's Objectives and Prior Work}
MMformer successfully achieved its research objectives, yielding significant results in practical applications. Central to its success is the ATMA mechanism, which enables the model to "learn to learn" via meta-learning, facilitating rapid adaptation to data distribution shifts and novel forecasting tasks. It effectively addresses challenges related to computational load and data heterogeneity across multiple regions' time series analysis. Furthermore, the application of MC Dropout not only enhances model robustness but also provides crucial uncertainty quantification, addressing the lack of uncertainty quantification challenge, a significant advance for high-stakes environmental decision-making. Quantitative comparisons with leading models, including iTransformer, TimesNet, PatchTST, and standard Transformer, demonstrate superior performance across air quality, climate datasets, and PEMS datasets. Specifically, MMformer achieved at least a 76.598\% reduction in MSE, a 41.696\% reduction in MAE, and a 40.097\% reduction in MAPE on the air quality and climate datasets. It also achieved SOTA results on PEMS03, PEMS04, and PEMS08. These findings validate MMformer's approach in overcoming the common limitations of existing deep learning models on spatio-temporal environmental data, particularly in capturing complex dynamics, inter-variable dependencies, and handling noise and uncertainty.

\subsection{Managerial Insights and Practical Implications for Policymakers}
MMformer's enhanced forecast accuracy and reliability have significant management implications for policymakers. Proven by air quality, climate, and PEMS predictions, the model aids decision-makers in understanding environmental trends, developing more effective pollution control strategies, and optimizing resource allocation. The uncertainty quantification provided is crucial for risk assessment and the formulation of responsive measures. For instance, identifying high-risk areas allows for targeted resource deployment, mitigating environmental risks, and improving public health outcomes.

\subsection{Theoretical Insights for Researchers}
MMformer introduces novel perspectives for deep learning applications across multiple regions for MTS prediction. The integration of the ATMA mechanism and MC Dropout offers a generalizable methodology, enhancing model adaptability, robustness, and uncertainty quantification across diverse datasets. This study underscores the critical importance of considering prediction uncertainty in spatiotemporal data analysis. These insights are poised to drive further advancements in deep learning models for complex, dynamic, and uncertain environments across multi-regional time series.

\subsection{Limitations}
While MMformer has demonstrated significant advancements in multi-regional MTS prediction, certain limitations persist. Firstly, for complex single-city datasets with sparse features, the meta-learning advantage of ATMA is less pronounced, resulting in low precision. Secondly, the integration of MC Dropout for uncertainty quantification may lead to a trade-off with predictive accuracy. Thirdly, despite ATMA's adaptive capabilities, the model still faces challenges in adapting to highly non-stationary data distributions across multiple locations. Finally, the model exhibits a degree of 'black box' behavior, limiting deeper interpretability. The following section will summarize these findings and outline future research directions.

\section{Conclusion and Future Work}
This study introduces MMformer, a model designed to address key challenges in multi-regional MTS forecasting, including accuracy, adaptability, robustness to complex data, and reliable uncertainty quantification. MMformer achieves significant performance improvements on real-world environmental datasets (air quality, climate) and general MTS data (PEMS) by integrating the ATMA mechanism and MC Dropout. The core innovation lies in the synergistic integration of these components and their effective application in environmental MTS forecasting, providing enhanced adaptability and uncertainty insights compared to individual or simpler combinations.

Comprehensive experiments and ablation studies demonstrate that MMformer achieves SOTA or near-SOTA results compared to established models (iTransformer, TimesNet, PatchTST, Transformer) across air quality, climate, and PEMS datasets. Specifically, on air quality and climate datasets, MMformer shows significant percentage reductions in MSE, MAE, and MAPE compared to baseline models (e.g., at least 68.182\% MSE reduction, 36.641\% MAE reduction, 35.343\% MAPE reduction), showcasing superior error minimization and data fitting. PEMS dataset experiments validate MMformer's generalizability, with best performance on multiple datasets. Ablation study results consistently highlight the critical role of the meta-learning component and MC Dropout in enhancing MMformer's overall performance, accuracy, and robustness. Although iTransformer slightly outperforms on the MAPE metric for the PEMS07 dataset, MMformer consistently demonstrates strong predictive capabilities and robustness, especially when handling complex, heterogeneous, and uncertain environmental MTS data.

Future work will focus on addressing limitations identified in this study and expanding MMformer's capabilities and applications: (1) Enhancing computational efficiency and scalability by developing more efficient, lightweight architectures and applying model distillation and sparse attention mechanisms for real-time deployment. (2) Improving handling of extreme events and outliers by integrating self-supervised learning and causal inference to improve the detection of extreme and rare events and improve model adaptability and generalization in boundary cases. (3) Increasing model interpretability by exploring explainable time series deep learning models, such as incorporating visual attention mechanisms and causal tracing techniques, to enhance model transparency and user trust. (4) Extending the model's applicability by expanding MMformer's transfer capabilities to handle cross-domain and multi-modal data, and integrating online and active learning to support dynamic, real-time decision-making.

\section*{Declaration of Competing Interest}
The authors declare that they have no known competing financial interests or personal relationships that could have appeared to influence the work reported in this study.

\section*{Acknowledgement}
We sincerely thank the Climatic Data Centre, part of the National Meteorological Information Centre (CMA Meteorological Data Centre), for their invaluable assistance and cooperation in providing us with the meteorological data used in this study.

 \bibliographystyle{elsarticle-num} 
 \bibliography{elsarticle-template-num}




\appendix

\clearpage 

\section{Abbreviations}
\begin{longtable}{@{}lll@{}}
        \toprule 
        \textbf{Abbreviations} & \textbf{Full name} \\
        \midrule 
        MTS & Multivariate Time Series \\	
        DL & Deep learning \\
        ATMA & Adaptive Transferable Multi-head Attention\\	
        MC Dropout & Monte Carlo Dropout \\
        MLPs & Multi-Layer Perceptrons \\
        FFN & Feed-forward networks \\
        SOTA & State-of-the-art \\
        AQI & Air Quality Index \\
        PM2.5 & Fine particulate matter less than or equal to 2.5 microns in diameter \\
        PM10 & Inhalable particulate matter with a diameter less than or equal to 10 microns \\
        SO$_2$ & Sulfur dioxide \\
        NO$_2$	& Nitrogen dioxide \\
        CO	& Carbon monoxide \\
        O$_3$	& Ozone \\
        GLMs & Generalized Linear Models \\
        SARIMA & Seasonal Autoregressive Integrated Moving Average \\
        ARIMA & Autoregressive Integrated Moving Average \\
        MAML & Model-Agnostic Meta-Learning \\
        RNNs & Recurrent Neural Networks \\
        LSTM & Long Short-Term Memory \\
        GRU & Gated Recurrent Unit \\
        SVMs & Support Vector Machines \\
        MAE & Mean Absolute Error \\
        MSE & Mean Squared Error \\
        MAPE & Mean Absolute Percentage Error \\
        seq\_len & Sequence length \\
        pred\_len & Prediction length \\
        IQR	& Interquartile range \\
        \bottomrule 
\end{longtable}
\end{sloppypar}
\end{document}